\documentclass[aps,pre,preprint,groupedaddress,amsmath,amssymb]{revtex4-1}

\usepackage{graphicx}
\usepackage{dcolumn}
\usepackage{bm}
\usepackage{subcaption}  

\begin{document}


\title{Flexible Dynamics of Two Quorum Sensing Coupled Repressilators}


\author{Edward H. Hellen}
\email[]{ehhellen@uncg.edu}
\affiliation{Department of Physics \& Astronomy, University of North Carolina Greensboro, Greensboro, NC USA}

\author{Evgeny Volkov}
\affiliation{Department of Theoretical Physics, Lebedev Physical Institute, Moscow, Russia}

\date{\today}

\begin{abstract}
Genetic oscillators play important roles in cell life regulation. The regulatory efficiency usually depends strongly on the emergence of stable collective dynamic modes, which requires designing the interactions between genetic networks. We investigate the dynamics of two identical synthetic genetic repressilators coupled by an additional plasmid which implements quorum sensing (QS) in each network thereby supporting global coupling. In a basic genetic ring oscillator network in which three genes inhibit each other in unidirectional manner, QS stimulates the transcriptional activity of chosen genes providing for competition between inhibitory and stimulatory activities localized in those genes. The ``promoter strength'', the Hill cooperativity coefficient of transcription repression, and the coupling strength, i.e., parameters controlling the basic rates of genetic reactions, were chosen for extensive bifurcation analysis. The results are presented as a map of dynamic regimes. We found that the remarkable multistability of the anti-phase limit cycle and stable homogeneous and inhomogeneous steady states exists over broad ranges of control parameters. We studied the anti-phase limit cycle stability and the evolution of irregular oscillatory regimes in the parameter areas where the anti-phase cycle loses stability. In these regions we observed developing complex oscillations, collective chaos, and multistability between regular limit cycles and complex oscillations over uncommonly large intervals of coupling strength. QS-coupling stimulates the appearance of intrachaotic periodic windows with spatially symmetric and asymmetric partial limit cycles which, in turn, change the type of chaos from a simple anti-phase character into chaos composed of pieces of the trajectories having alternating polarity. The very rich dynamics discovered in the system of two identical simple ring oscillators may serve as a possible background for biological phenotypic diversification, as well as a stimulator to search for similar coupling in oscillator arrays in other areas of nature, e.g. in neurobiology, ecology, climatology etc.
\end{abstract}


\maketitle


\section{Introduction}
Synthetic genetic networks provide researchers with the opportunity of designing biologically based circuitry for accomplishing specific functions. In principle, such circuits can be incorporated into natural cellular machinery or used in an entirely synthetic environment.  Oscillators are essential motifs for circuit design \cite{tyson2010}.  The repressilator is a synthetic genetic oscillator (GO) in the form of a ring of three genes sequentially inhibiting one another's transcription. The GO has been inserted experimentally into \textit{E. Coli} \cite{elowitz2000} and extensively studied theoretically via deterministic and stochastic approaches \cite{loinger2007,buse2010,strelkowa2010}.  

Coupling of individual GOs is important in coordinated activity of GO populations. Bacterial Quorum Sensing (QS) \cite{waters2005}, which provides for cell-cell communications in bacterial populations by fast diffusion of small specific molecules (autoinducers), is a natural candidate for the role of being a manager for synthetic genetic network dynamics. This idea was effective, for instance, in constructing several synthetic multicellular systems like the ecological predator-prey \cite{balagadde2008}, population control \cite{you2004}, and other models (see Refs. in recent review \cite{hennig2015}).  

QS-coupling has been used for GO synchronization in different mathematical models \cite{mcmillen2002,garcia2004,kuznetsov2004}, as well as in experimental demonstration of a multicellular clock \cite{danino2010}. Recently, the QS system has been used to construct a synthetic microbial consortium with population-level oscillations \cite{chen2015}.  Although autoinducers freely diffuse between cells with GOs inside, this type of coupling as a whole cannot be reduced to simple linear diffusive coupling which still dominates in studies of coupled oscillators. An autoinducer is not a required element of a GO, its production may be controlled by one gene of the GO networks but its target is transcription regulation of a different gene in the same or neighboring cells. Being physically diffusive, this coupling mechanism is, however, difficult to present as just one classical diffusive term in the ODE system describing the genetic circuits. To write down the coupling term mathematically it is necessary to consistently consider the particular reaction scheme. This way to draw the coupling term seems quite natural for biochemical and genetic networks in which communication is a cumulative multistep process which includes regulative production, diffusion, binding to targets, and other important metabolic steps.    

The first attempt to demonstrate theoretically the possibility of in-phase synchronization of repressilators was successfully realized \cite{garcia2004} and checked in a simple electronic model \cite{wagemakers2006}, but not all collective regimes were detected within the limits of the coupling design considered. Later publication demonstrated that the use of limited parameter space masked other dynamical regimes. The anti-phase limit cycle (APLC) and complex periodic regimes were revealed when stronger  repression in transcriptional regulation and greater differences in time scales of mRNA and transcription factor kinetics were taken into account \cite{potapov2011}.

In the next version \cite{ullner2007,ullner2008} of QS-dependent cell-cell interaction, identical repressilators were coupled using a modification of the additional plasmid for the QS mechanism. The modification provided phase-repulsive interaction between oscillators which leads to a rich set of stable attractors:  periodic regular anti-phase limit cycle (APLC – time series are shifted by half-period), stable homogeneous steady state (HSS - the identical values of the same name variables in each oscillator), inhomogeneous steady states (IHSS - different values of variables), inhomogeneous limit cycle (IHLC) emerging from IHSS, and chaotic regime which appears via torus bifurcation of the AP regime.  Some regimes can coexist over broad intervals of model parameters, which opens a possibility for switching between attractors in the presence of extrinsic/intrinsic noise.  Specific examples of all these regimes will be presented in Results section. 

An important problem in the studies of gene expression regulation is very limited knowledge of the parameter values. The transcription regulation is typically described by Hill function, $\alpha/(1+x^n)$, where the main parameters are the maximum transcription rate ($\alpha$) and the degree of cooperativity ($n$) of transcription factor ($x$) binding to promoter.  Previous publications \cite{ullner2007,ullner2008} concentrated on the dynamics with small Hill repression $n$ and limited values for transcription rates, time scales ratios for mRNA and proteins kinetics, and QS signaling molecule (autoinducer) activity as transcription activator. The goals of this paper are to significantly extend the main parameter areas within the limits of one model of QS-coupled identical or near identical repressilators \cite{ullner2008} to detect  new dynamic behavior(s), to present the coarse-grained structure and content of the phase diagram (the map of regimes), and to investigate the robustness of multistability with respect to parameter values. We will use a reduced version of the mathematical model, as well as its electronic circuit model \cite{hellen2011,hellen2016} adapted for two coupled 4-dim repressilators. The effectiveness of this analog version has been checked recently in the study of a single repressilator with QS feedback \cite{hellen2013b}. Use of the inherently different numerical and electronic models provides a test of the robustness of the dynamics. We consider a wide range of control parameters, even going beyond the characteristic limits known for them today. The results obtained using the extended ranges may be viewed as predictions of the model for the future, when these extended limits become accessible with the fast development of synthetic genetics. The dynamical results produced by the QS-coupled oscillators used here may inspire application to other systems beyond synthetic gene circuits, such as electronic ones (see below) or chemical systems of interacting water droplets bearing oscillating BZ reactions coupled via diffusion in selective oil environments \cite{toiya2008}. 
       
We find that for a small degree of repression of transcription factor binding ($n<3$) anti-phase LC is a single homogeneous periodic attractor, and that this attractor loses stability via torus bifurcation if $n$ increases ($n=3$) and gets replaced by complex/chaotic regimes over a large interval of coupling strengths ($Q$). This $Q$-continuation branch of the AP limit cycle and chaos coexists with stable homogeneous and inhomogeneous SS and IHLC providing a rich landscape of attractors in multi-parameter space. 

The sizes of parameter's intervals occupied by the complex regimes are unusually large and include many periodic windows which contain both spatially symmetrical and asymmetrical stable limit cycles. The presence of asymmetrical limit cycles results in the appearance of chaotic trajectories built from pieces with randomly alternating ``polarity''. This type of chaotic time series differs from that of simple anti-phase chaos observed in parameter regions devoid of asymmetrical limit cycles.  
   
The extension of the ranges for parameter analysis also reveals a new type of multistability if changes in the maximum transcription rate ($\alpha$) are taken into account. For a value of the Hill coefficient near 3 a range of $\alpha$ gives rise to the coexistence of the complex nonperiodic oscillations and the anti-phase limit cycle with 5 return times in one period. Again, the range of this hysteresis covers large intervals of the coupling strengths and $\alpha$.

A further increase in transcription cooperativity up to $n=4$ leads to the recovery of stability of the in-phase limit cycle in addition to stable APLC. In-phase limit cycle is not a dominant regime but its appearance seems important to complete the entire dynamic picture demonstrated by the two coupled identical repressilators.      

\section{Numerical and Electrical Models}
\begin{figure}
\includegraphics[width=0.75\linewidth]{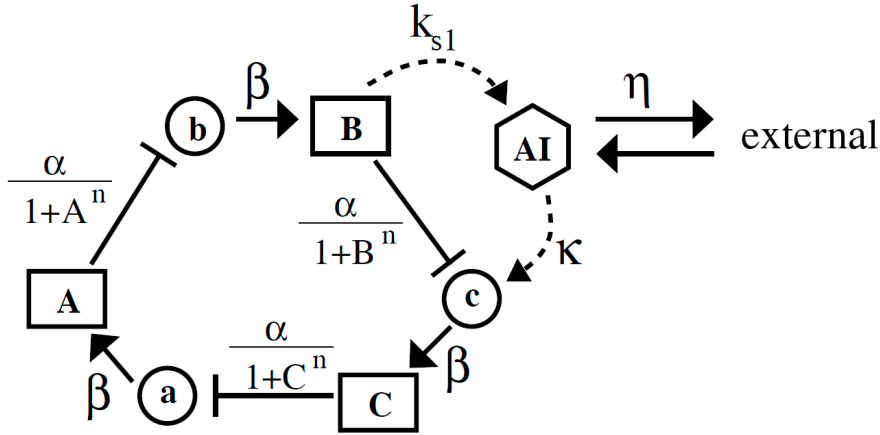}%
\caption{\label{RepQs}A repressilator genetic network with QS feedback.  Lower case (a, b, c) are mRNA and upper case (A, B, C) are expressed protein repressors. AI is the auto-inducer molecule which diffuses through the cell membrane.}%
\end{figure}

We investigate the dynamics of two repressilators interacting via repressive QS coupling as used previously \cite{ullner2007,ullner2008}.  Figure \ref{RepQs} shows a single repressilator coupled via QS to the external medium. The three genes in the loop produce mRNA ($a, b, c$) and proteins ($A, B, C$), and they impose Hill function inhibition on each other in cyclic order by the preceding gene. The QS feedback is maintained by the AI produced (rate $k_{S1}$) by the protein B while the AI communicates with the external environment and activates (rate $\kappa$ in combination with Michaelis function) production of mRNA for protein C, which, in turn, reduces the concentration of protein A resulting in activation of protein B production. In this way the protein B plays a dual role of direct inhibition of protein C synthesis and AI-dependent activation of protein C synthesis, resulting in complex dynamics of the repressilator, even for just a single repressilator \cite{hellen2013b}.

The original models of a single repressilator \cite{elowitz2000,ullner2007} used re-scaled dimensionless quantities for rate constants and concentrations. We reduce the model for the case of fast mRNA kinetics ($(a, b, c)$ are assumed in steady state with their respective inhibitors $(C, A, B)$, so that $da/dt = db/dt = dc/dt \approx 0$). The resulting equations for the protein concentrations and AI concentration S are,
\begin{subequations}
\begin{align}
\frac{dA_i}{dt}=&\beta_1\left(-A_i+\frac{\alpha}{1+C_i^n}\right)\\
\frac{dB_i}{dt}=&\beta_2\left(-B_i+\frac{\alpha}{1+A_i^n}\right)\\
\label{eq_c}
\frac{dC_i}{dt}=&\beta_3\left(-C_i+\frac{\alpha}{1+B_i^n}+\frac{\kappa S_i}{1+S_i}\right)\\
\frac{dS_i}{dt}=&-k_{S0}S_i+k_{S1}B_i-\eta\left(S_i-S_{ext}\right)
\end{align}
\label{ode}
\end{subequations}
where $i=1,2$ for the two repressilators, $\beta_j (j=1,2,3)$ are the ratios of protein decay rate to mRNA decay rate, $\alpha$ accounts for the maximum transcription rate in the absence of an inhibitor, $n$ is the Hill coefficient for inhibition, and $k_{S0}$ is the ratio of AI decay rate to mRNA decay rate. The diffusion coefficient $\eta$ depends on the permeability of the membrane to the AI molecule.  The concentration of AI in the external medium is $S_{ext}$ and is determined according to quasi-steady-state approximation by AI produced by both repressilators ($S_1$ and $S_2$), and a dilution factor $Q$. 
\begin{equation}
S_{ext}=Q\frac{S_1+S_2}{2}
\end{equation}
Numerical simulations are performed with XPPAUT \cite{ermentrout2002} and by direct integration with 4-th order Runge-Kutta solver.  We choose parameter values similar to ones used previously shown to be experimentally reasonable taking into account realistic biochemical rates and binding affinities \cite{ullner2008}. Here we use $\beta_1 = 0.5, \beta_2 = \beta_3 =0.1, n = 2.8$ to $4, \alpha \approx 200, k_{S0} = 1, k_{S1} = 0.01$, and $\eta = 2$.  

\begin{figure}
\begin{subfigure}{0.49\textwidth}
\includegraphics[width=0.85\linewidth]{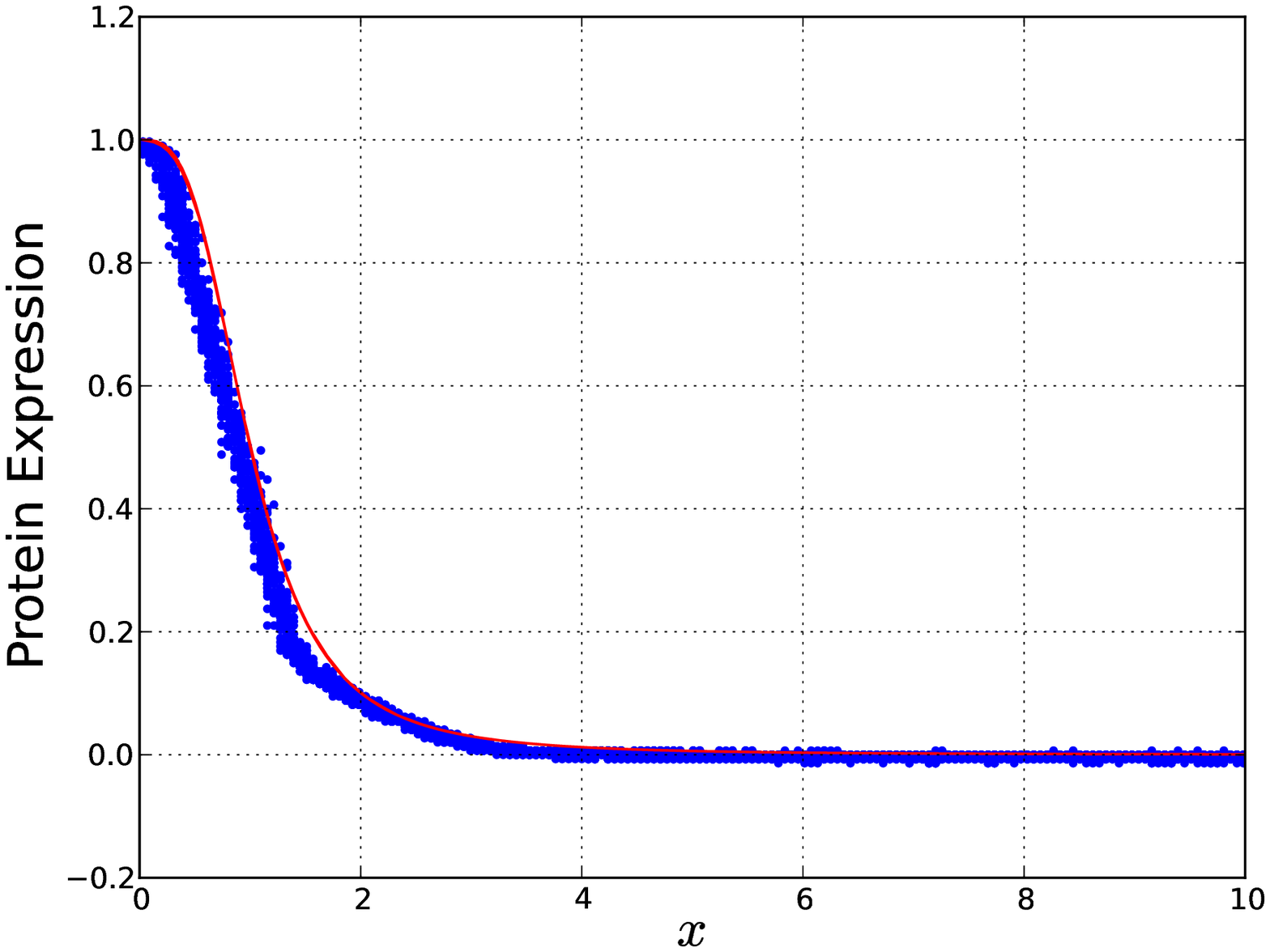}
\caption{\label{eHill} Hill inhibition for $n=3.17$, $\alpha=218$.}
\end{subfigure}\
\begin{subfigure}{0.49\textwidth}
\includegraphics[width=0.85\linewidth]{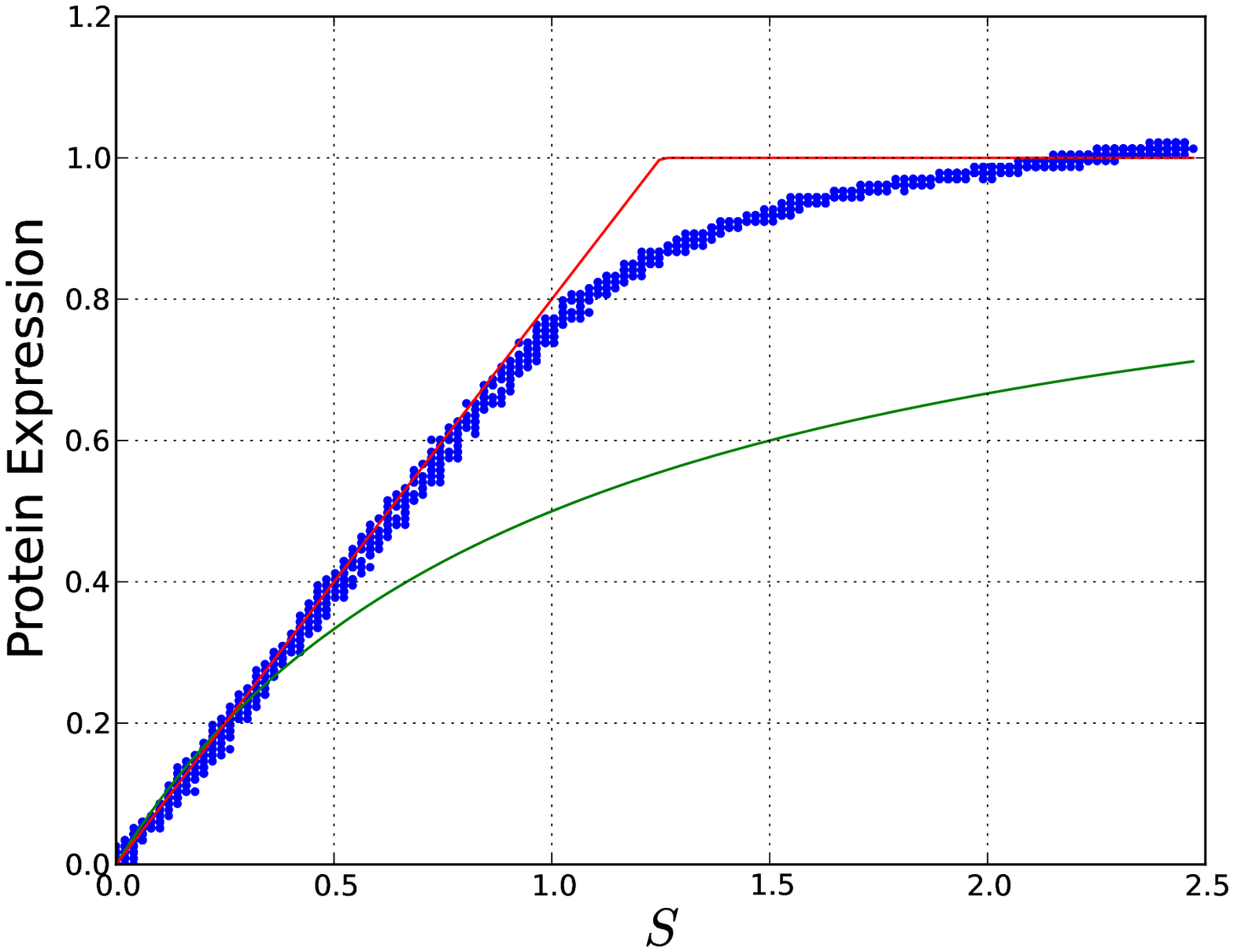}
\caption{\label{QS-activation}QS-activation for $\kappa=21$, $\alpha=135$.}
\end{subfigure}
\caption{\label{circuit}Comparison of circuit behavior (blue) and mathematical model (red) for (a) inhibition and (b) activation. Panel (b) also shows $S/(1+S)$ as green line.}
\end{figure}

We also implement the model Eq.\ \eqref{ode} in electronic circuits as described previously \cite{hellen2011,hellen2013b}, with additional performance improvements \cite{hellen2016}.  By its very nature the circuit has inherent parameter mismatch and noise, and is free from any numerical artifacts.  It provides experimental results which complement numerical simulations. Model parameter values are set in the circuit by resister values, capacitor values, and reference voltages as described in Supplemental Materials (SM).
The hyperbolic dependence $S/(1+S)$ in Eq.\ \eqref{eq_c} is replaced in the circuit by the linear-piecewise-continuous behavior min($0.8S,1$) as described previously \cite{hellen2013b}.  Therefore Eq.\ \eqref{eq_c} in simulations is replaced by
\begin{equation}
\frac{dC_i}{dt}=\beta_3\left(-C_i+\frac{\alpha}{1+B_i^n}+\kappa \text{min}\left( 0.8S_i,1\right)\right).
\label{min-S}
\end{equation}
An important goal of the circuit design is to reproduce from Eqs.\ \eqref{ode} and \eqref{min-S}  the Hill function inhibition $1/(1+x^n)$, and the QS-activation min($0.8S,1$). Figure \ref{circuit} shows the resulting comparison of measured circuit performance and mathematical model for inhibition and activation. 

\section{Results}
\subsection{Low Repression: $n=2.8$}
We start with the demonstration of the general structure of the phase diagram (map of regimes) for two coupled 4-dim repressilators (Eq.\ \eqref{ode}) with parameters similar to those used previously \cite{ullner2008} in a study with non-reduced coupled 7-dim repressilators.  We choose protein B as the dynamical variable of interest. 

\begin{figure}
\begin{subfigure}{0.49\textwidth}
\centering
\includegraphics[width=0.85\linewidth]{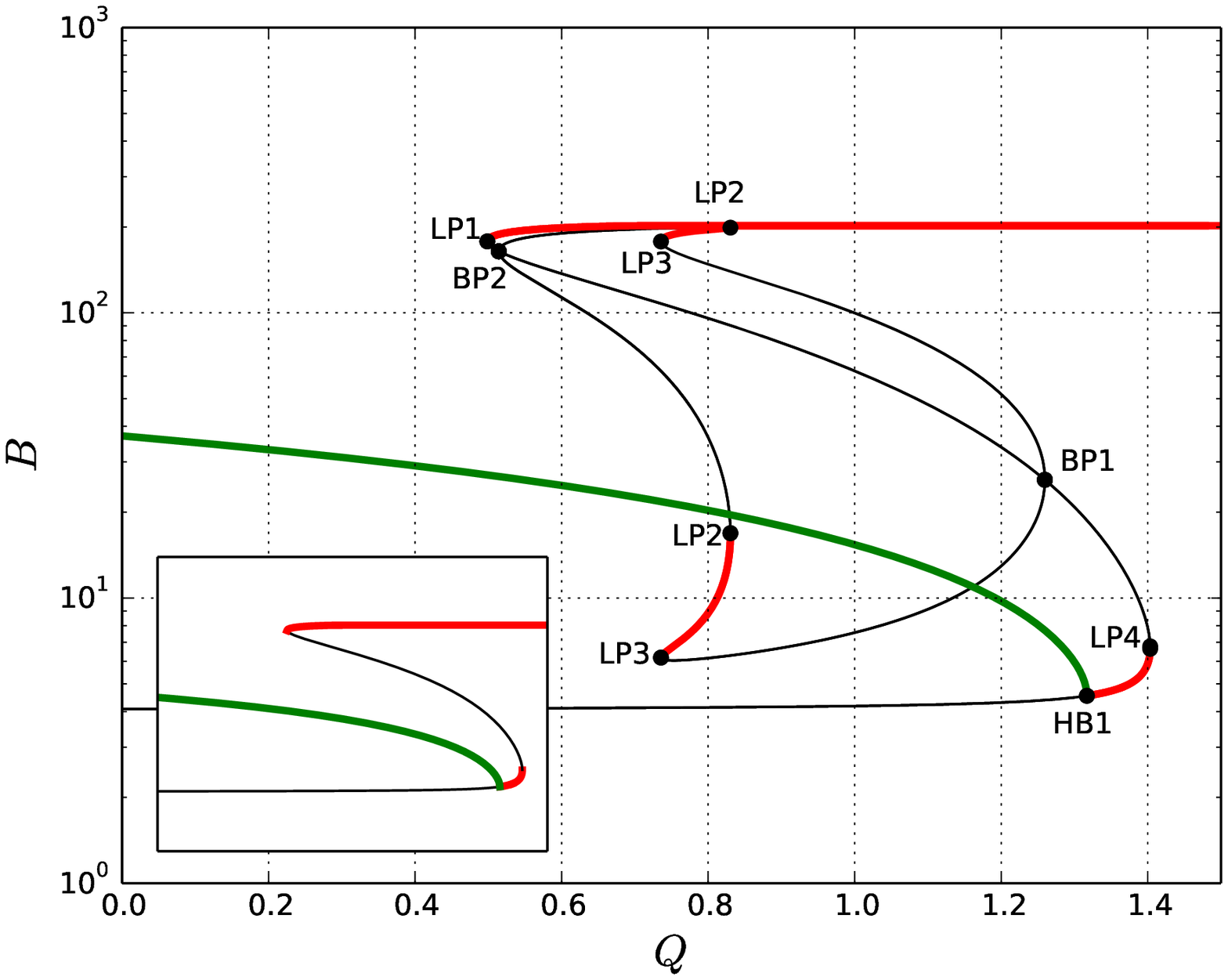}
\caption{\label{n2.8k12}$\kappa=12$. Inset shows just the homogeneous branch.}  
\end{subfigure}\
\begin{subfigure}{0.49\textwidth}
\centering
\includegraphics[width=0.85\linewidth]{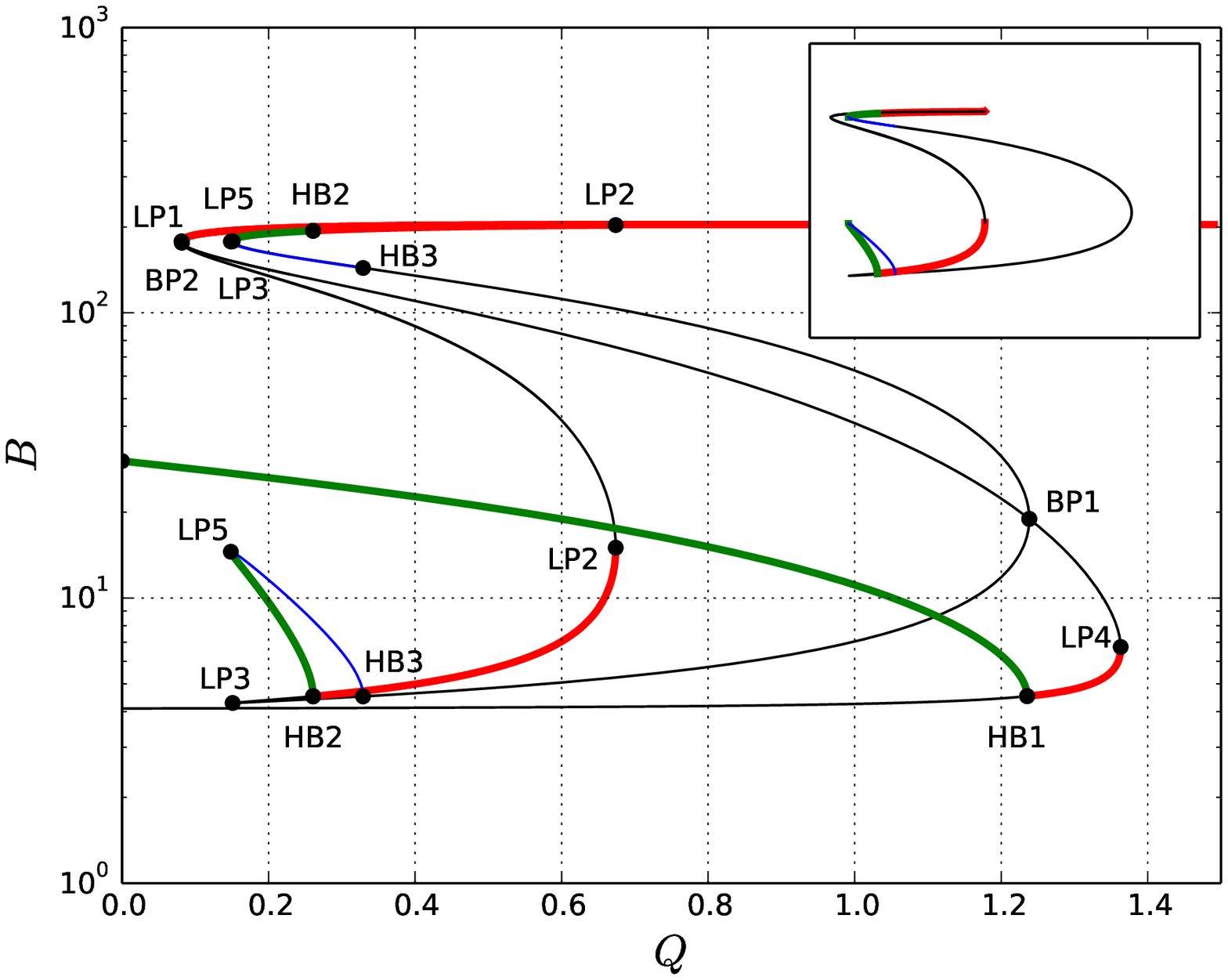}
\caption{\label{n2.8k17}$\kappa=17$.  Inset shows just the inhomogeneous branch.}  
\end{subfigure}
\caption{\label{qbifs-2.8}Numerical XPP/AUTO continuations showing the values of proteins $B_i$ for steady states and the amplitudes of oscillations for limit cycles. HB--Andronov-Hopf bifurcation, LP--limit point, BP--branch point (symmetry breaking bifurcation). Solid thick (thin) lines are stable (unstable) attractors.  $n=2.8$, $\alpha=204$. Stable (red) and unstable (black) steady state, stable (green) and unstable (blue) limit cycle. Two red lines between LP3 and LP2 correspond to values $B_1$ and $B_2$ for IHSS. Panel (b) two green lines between LP5 and HB2 are amplitudes $B_1$ and $B_2$ for stable IHLC, and two blue lines between LP5 and HB3 are for unstable IHLC.}
\end{figure}

Figures \ref{n2.8k12} and \ref{n2.8k17} show bifurcation diagrams with Hill coefficient repression $n = 2.8$ for two different activation rates, $\kappa = 12$ and 17.  In these figures the coupling strength $Q$ is varied and the protein B maxima of resulting dynamical regimes are plotted: stable steady state (red), unstable steady state (black), stable LC (green), unstable LC (blue).  There are homogeneous dynamics in which both repressilators have the same maximum value and inhomogeneous (IH) dynamics where they have different maximum values. For clarity, the homogeneous branch is shown alone in the inset in Fig.\ \ref{n2.8k12} and the IH branch is shown alone in the inset in Fig.\ \ref{n2.8k17}.  The branch points (BP1 and BP2 in Figs.\ \ref{n2.8k12} and \ref{n2.8k17}) are where the homogeneous and IH steady-state branches intersect. The coupling strength is varied from 0 to 1.5 in order to show the entire structure of the system, although values of $Q > 1$ are not accessible in the conventional biological system.  

\begin{figure}
\begin{subfigure}{0.49\textwidth}
\centering
\includegraphics[width=0.85\linewidth]{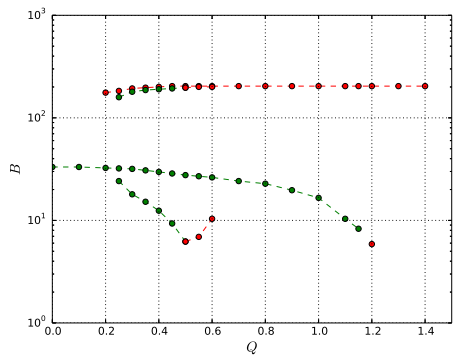}
\caption{\label{n2.8circ}Circuit Q-continuation of measured protein $B_i$ voltages for steady states (red) and voltage amplitudes for limit cycles (green).}  
\end{subfigure}\  
\begin{subfigure}{0.49\textwidth}
\centering
\includegraphics[width=0.85\linewidth]{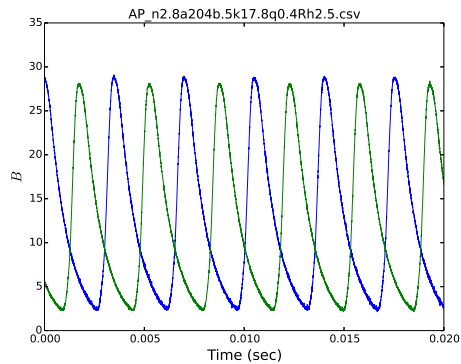}
\caption{\label{n2.8APcirc}Circuit protein $B_i$ voltages for APLC. $Q=0.4$.}  
\end{subfigure}
\caption{Circuit measurement results. $n=2.8,\alpha=204, \kappa=17.8$.}
\end{figure}

Figure \ref{n2.8k12} for activation strength $\kappa = 12$ shows a variety of dynamical behaviors.  The main features are: the stabilization of the high-B-HSS for large coupling strength ($Q > 0.5$); APLC(green) (see time series in Fig.\ \ref{n2.8APcirc}) which is the only stable dynamical behavior for small coupling strength (Q below LP1) and which remains stable out to large coupling strength at HB1 where the AP converts to a low-B-HSS; and the broken symmetry bifurcations (BP1 and BP2) where IH solutions arise including the stable IHSS between LP2 and LP3. There is coexistence of HSS, IHSS, and AP for Q between LP2 and LP3.  This narrow Q-range of 3-state multistability is embedded within the broader range of 2-state multistability.  Figure \ref{n2.8k12} displays versatile multistability controlled by coupling strength Q.

Figures \ref{n2.8k12} and \ref{n2.8k17} show that increasing the activation strength $\kappa$ from 12 to 17 has two qualitative effects.  First, the increase produces an IHLC (see time series in Fig.\ 5a) and its associated HB2, which appears where the IHSS becomes unstable between LP2 and LP3.  The IHLC extends from HB2 to LP5.  Second, the coupling strength Q-range for multistability is shifted and increased.  For $\kappa = 12$, AP is the only stable behavior for $Q <0.5$, whereas for $\kappa = 17$ the regime of multistability extends down to $Q=0.1$, and the Q-range for 3-state multistability has increased about 4-fold. We note that the LP1 and BP2 points in Fig.\ \ref{n2.8k17} are too close together to be resolved on the figure, as are LP3 and LP5 for the upper piece of the IHLC.  Also, the LP5 of the IHLC extends to a slightly lower Q-value than does the LP3 of the unstable portion of the IH branch. The sensitivity of the dynamics in Fig.\ \ref{qbifs-2.8} was investigated by varying parameters $\alpha$ and $\beta_1$.  Figures SM3 and SM4 show that these dynamics are not restricted to a narrow parameter range.   

Circuit results for $n=2.8, \kappa=17.8$ generally confirm the predictions in Fig.\ \ref{n2.8k17}.  Figure \ref{n2.8circ} shows a bifurcation diagram constructed from measured voltage amplitudes for the various dynamics.  The APLC-branch is stable to $Q=1.2$ where it converts to low-B-HSS, IHLC occurs for $0.25<Q<0.5$, IHSS occurs in $0.5<Q<0.6$, and HSS exists for $Q > 0.2$.  The low-B-HSS exists over a Q-range narrower than the data resolution and therefore appears as a single point at $Q=1.2$.  Figures \ref{n2.8k12}, \ref{n2.8k17}, and \ref{n2.8circ} show that for $n=2.8$ APLC is the single homogeneous periodic attractor.

\begin{figure}
\begin{subfigure}{0.495\textwidth}
\centering
\includegraphics[width=0.5\linewidth]{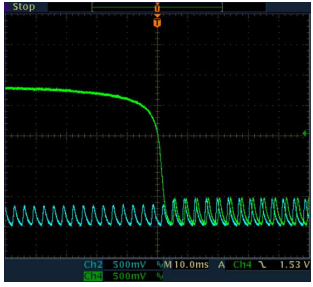}
\caption{\label{screenshota}IHLC to APLC for decreasing $Q$ at 0.25.}  
\end{subfigure}
\begin{subfigure}{0.495\textwidth}
\centering
\includegraphics[width=0.5\linewidth]{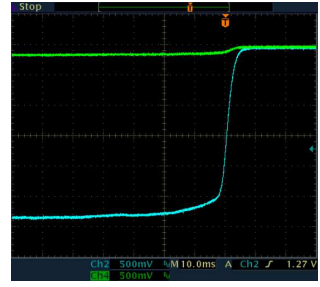}
\caption{\label{screenshotb}IHSS to HSS for increasing $Q$ at 0.6.}  
\end{subfigure}
\\
\begin{subfigure}{0.495\textwidth}
\centering
\includegraphics[width=0.5\linewidth]{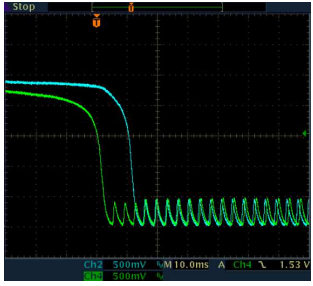}
\caption{\label{screenshotc}HSS to APLC for decreasing $Q$ at 0.15.}  
\end{subfigure}
\begin{subfigure}{0.495\textwidth}
\centering
\includegraphics[width=0.5\linewidth]{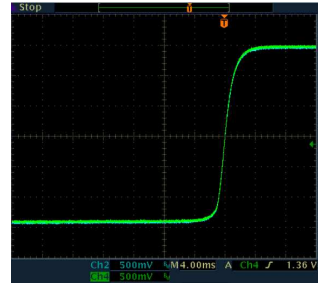}
\caption{\label{screenshotd}Low-B-HSS to HSS for increasing $Q$ at 1.2.}  
\end{subfigure}
\caption{Oscilloscope screenshots of transitions between stable dynamics in circuits. $n=2.8, \alpha=204, \kappa=17.8$.}
\label{screenshots}
\end{figure}

Measured time series of the dynamics represented in Fig.\ \ref{n2.8k17} are shown in Figs.\ \ref{n2.8APcirc} and \ref{screenshots}.  Digitized data showing the APLC oscillation for $Q = 0.4$ is shown in Fig.\ \ref{n2.8APcirc}.  Figure \ref{screenshots} screenshots show all four transitions corresponding to switching between stable branches in the bifurcation diagram as Q crosses limit points: Figure \ref{screenshota} shows IHLC to APLC transition for Q decreasing at 0.25 (LP5), Fig.\ \ref{screenshotb} shows IHSS to HSS for Q increasing at 0.6 (LP2), Fig.\ \ref{screenshotc} shows HSS to APLC for Q decreasing at 0.15 (LP1), and Fig.\ \ref{screenshotd} shows low-B-HSS to high-B-HSS for Q increasing at 1.2 (LP4).  In addition, the Fig.\ \ref{screenshots} screenshots show examples of the various dynamics prior to the transitions: IHLC in \ref{screenshota}, IHSS in \ref{screenshotb}, and low-B-HSS in \ref{screenshotd}. The high-B-HSS is clearly apparent after the transitions in \ref{screenshotb} and \ref{screenshotd}.   The difference in transition times in Fig.\ \ref{screenshotc} indicates that the HSS LP1 is sensitive to unavoidable differences between the two repressilator circuits.  We also point out that after the transitions in Figs.\ \ref{screenshota} and \ref{screenshotc} the phase-shifts are changing to the $180^{\circ}$ characteristic of APLC. 

We note that the Q-ranges for particular dynamics in Fig.\ \ref{n2.8circ} show some difference from the simulation in Fig.\ \ref{n2.8k17}.  However, overall agreement of the structure of dynamics is demonstrated.   Also, the circuit only finds stable dynamics (and transitions between stable states shown in Fig.\ \ref{screenshots}), and therefore Fig.\ \ref{n2.8circ} does not find the unstable branches (black lines) shown in the Fig.\ \ref{n2.8k17} simulations. 
 
\begin{figure}
\includegraphics[width=0.5\textwidth]{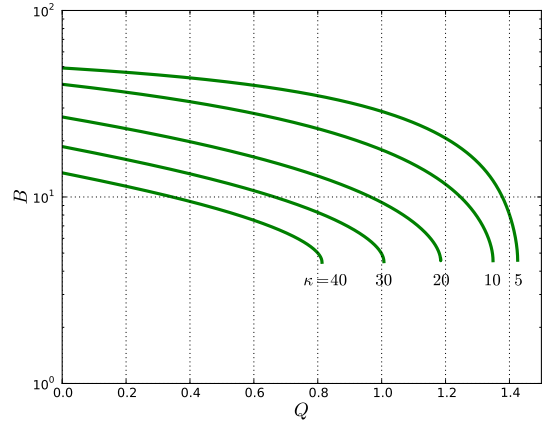}
\caption{\label{n2.8AP-ks}Numerical APLC-branch for various $\kappa$. $n=2.8, a=204$.}
\end{figure}

Figure \ref{n2.8AP-ks} shows simulations of the APLC-branch for $n=2.8, \alpha=204$ using various activation strengths $\kappa$.  As seen in Figs.\ \ref{n2.8k12} and \ref{n2.8k17} increasing $\kappa$ causes suppression of the APLC branch, but not instability.

The dual role of protein B both directly inhibiting production of C and indirectly activating C via QS is responsible for the expanded regime of high-B-HSS at larger activation $\kappa$ seen in Fig.\ \ref{n2.8k17}.  At low activation (small $\kappa$) and small coupling, each 3-gene ring essentially becomes an isolated repressilator, and therefore undergoes oscillations with no SS.  At large $\kappa$, B's activation of C overcomes its inhibition and thereby increases expression of protein C, which in-turn leads to inhibition of A, which results in high level expression of B creating a positive feedback for B, in contrast to the negative feedback in an isolated repressilator. 

The dynamics demonstrated in Figs.\ \ref{qbifs-2.8}-\ref{screenshots} both in simulations and in circuits are similar to those found previously for repressively coupled repressilators using similar parameter values \cite{ullner2008} and confirmed over large intervals of model parameters.  At low coupling strength Q the APLC is the only stable behavior, at higher Q IHLC and then IHSS coexist with both APLC and HSS.  In the  14-variable model \citep{ullner2008} with $n=2.6$ the APLC branch converts to chaotic behavior when Q reaches 0.6 whereas for the reduced 8-variable model presented here the AP branch remains stable.  However, as shown below, chaotic behavior is found at higher Hill coefficient of repression $n$.  

\subsection{Intermediate Repression: $n=3.0$}
\begin{figure}
\includegraphics[width=0.5\textwidth]{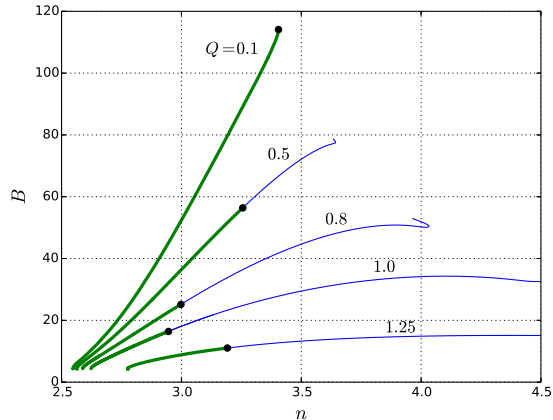}
\caption{\label{AP_n-contin}Stable (green) and unstable (blue) APLC-branch $n$-continuations. Black dots indicate TR (torus) bifurcations where APLC looses stability. $\kappa=15, \alpha=175$.}
\end{figure}

The structure of the phase diagrams in Fig.\ \ref{qbifs-2.8} for $n=2.8$ is very rich and it is interesting to understand its evolution and robustness under the change of such a basic parameter as the Hill repression $n$. A guide to the interesting $n$-values is provided by the $n$-continuation bifurcation diagram in Fig.\ \ref{AP_n-contin} for $\alpha=175, \kappa=15$, with coupling strengths $Q = (0.1, 0.5, 0.8, 1, 1.25)$.  Only the APLC branch is shown since this branch is the sole homogeneous periodic attractor. 

Interestingly, the $n$-range of stable APLC (green) is restricted between 2.5 and 3.5.  Decreasing $n$ below 2.8 will only reduce the amplitude of the stable APLC. Therefore we increase $n$, expecting the introduction of unstable regimes of APLC (blue in Fig.\ \ref{AP_n-contin}) at the larger coupling strengths.  Simulations also show that for $n>3.5$ large amplitude APLC is predicted for smaller $\kappa$ at small Q (essentially uncoupled repressilators), and small amplitude APLC is predicted for larger $\kappa$ at large Q ($\kappa$ suppresses and stabilizes the APLC, shown below).

\begin{figure}
\begin{subfigure}{0.495\textwidth}
\centering
\includegraphics[width=0.95\linewidth]{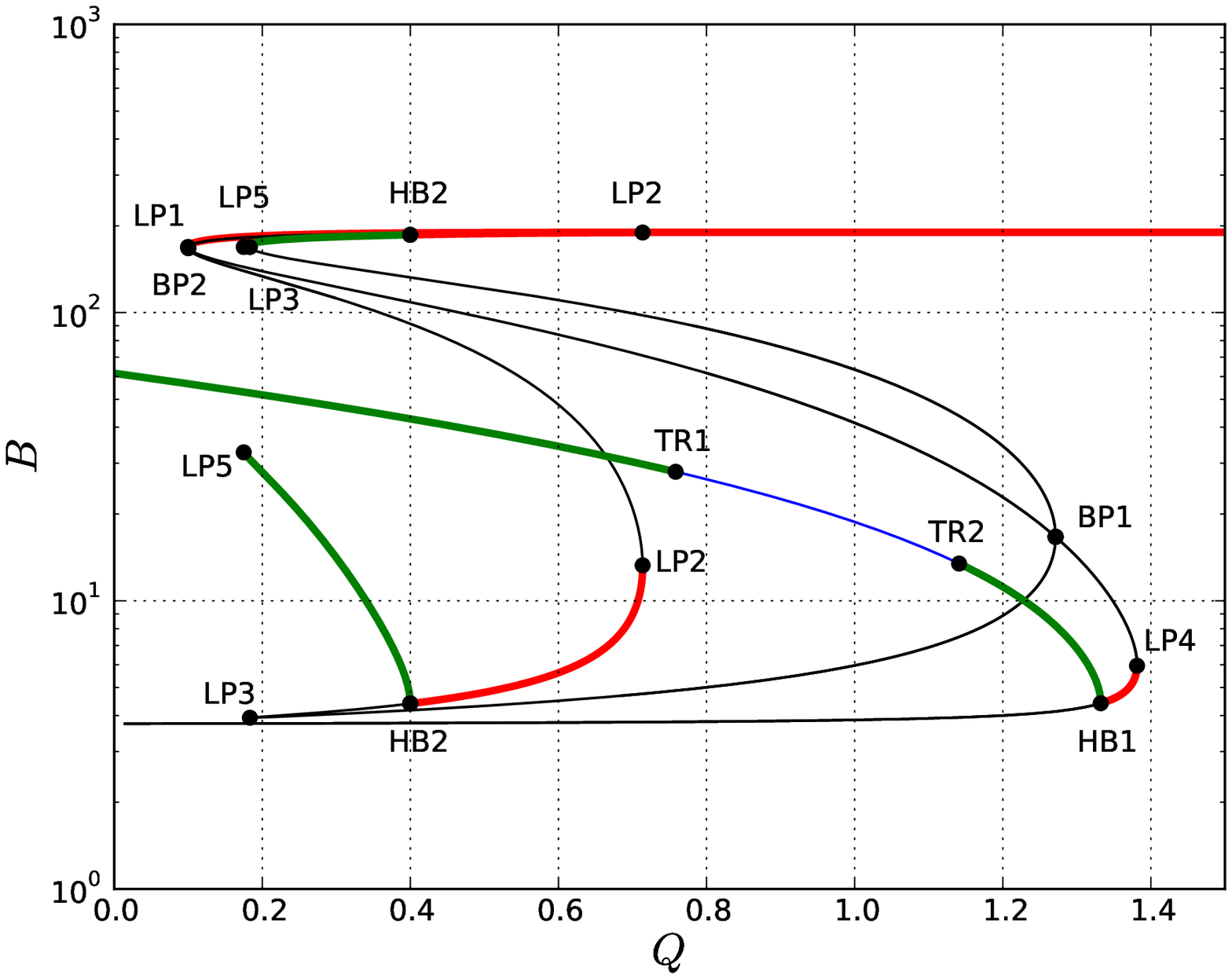}
\caption{\label{n3a190k15}All regimes. $\kappa=15$.}  
\end{subfigure}\
\begin{subfigure}{0.495\textwidth}
\centering
\includegraphics[width=0.95\linewidth]{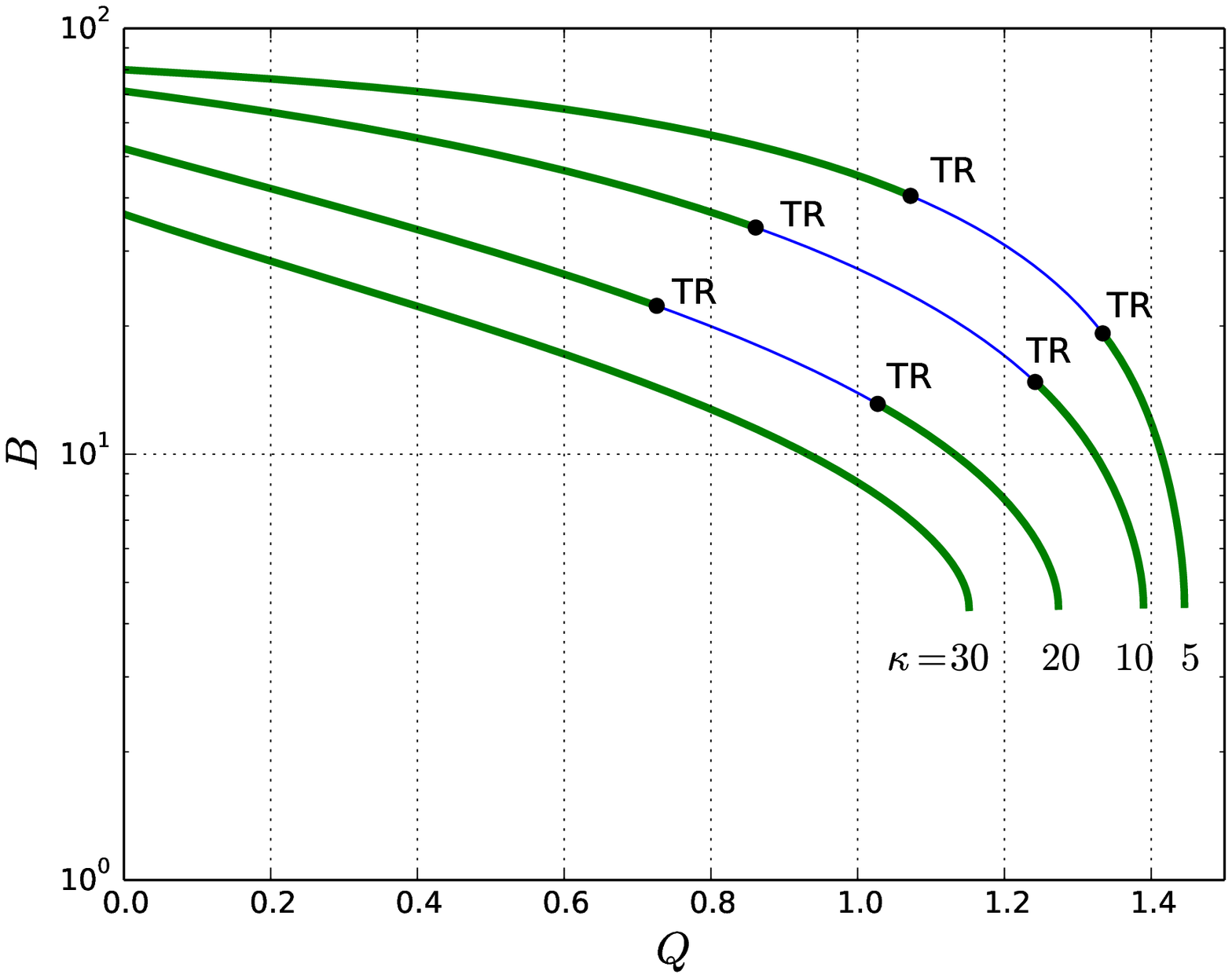}
\caption{\label{n3APs}APLC for various $\kappa$.}  
\end{subfigure}
\caption{Numerical Q-continuations as in Figs.\ \ref{qbifs-2.8} and \ref{n2.8AP-ks} for $n=3.0,\alpha=190$.}
\end{figure}

We begin by increasing repression to $n = 3.0$, with $\alpha=190$ and $\kappa = 15$. Figure \ref{n3a190k15} shows the simulated bifurcation diagram.  Most of the features are similar to those for the smaller repression case in Fig.\ \ref{n2.8k17}.  However increasing repression to $n=3.0$ has introduced the unstable region (blue) in the APLC branch between TR1 and TR2 as predicted in Fig.\ \ref{AP_n-contin}.  Simulations shown in Fig.\ \ref{n3APs} of the APLC-branch for various activation strengths demonstrate how increasing $\kappa$ suppresses and stabilizes the APLC-branch, as indicated by the continuous green branch for the largest activation $\kappa=30$.

\begin{figure}
\includegraphics[width=0.5\textwidth]{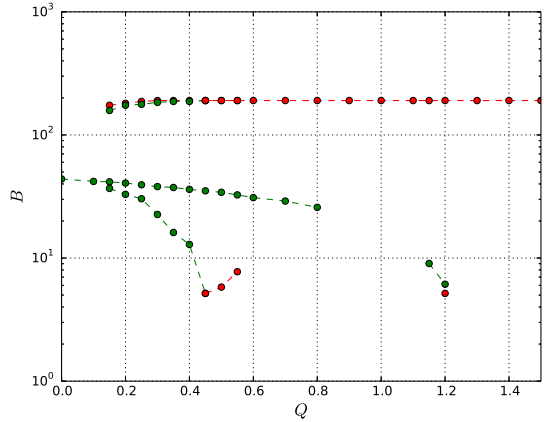}
\caption{\label{n3circ}Circuit Q-continuations of all regimes. $n=3.0, \alpha=190, \kappa=17.8$. Note the break in stability in the APLC-branch.}
\end{figure}

Figure \ref{n3circ} shows the circuit's measured bifurcation diagram for $n=3.0, \alpha=190, \kappa=17.8$.  Note the break in stability in the APLC branch as predicted by the simulations.  The similar structure of the dynamics in Figs.\ \ref{n3a190k15} and \ref{n3circ} demonstrates the agreement of circuit and simulations. Complex oscillations for $n= 3.0$ which occur between the TRs are discussed in section D on unstable APLC and chaos.  

\subsection{High Repression: $n>3.1$}
Further increase of Hill coefficient $n$ up to 3.2 does not result in qualitative changes in the structure of numerical and experimental bifurcation diagrams although the Q-intervals for IHLC and for chaos are enlarged. In the region of unstable APLC between the TRs a rich variety of behaviors are predicted for $n > 3$.  The nature of these behaviors is described below in the section D on unstable APLC and chaos.
\begin{figure}
\includegraphics[width=0.5\textwidth]{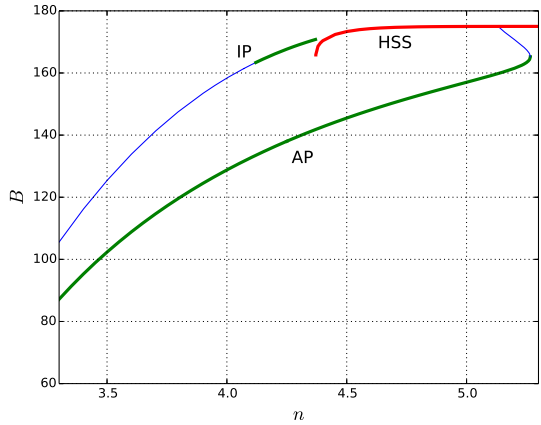}
\caption{\label{n-contink4}Numerical $n$-continuations of APLC, IPLC and HSS for $\alpha=175, \kappa=4, Q=0.8$.}
\end{figure}  

\begin{figure}
\includegraphics[width=0.4\textwidth]{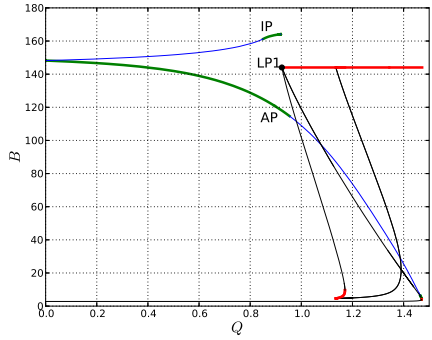}
\caption{\label{n4a175k4}Numerical Q-continuation for $n=4, \alpha=175, \kappa=4$ showing coexistence of IPLC and APLC.}
\end{figure}

\begin{figure}
\includegraphics[width=0.4\textwidth]{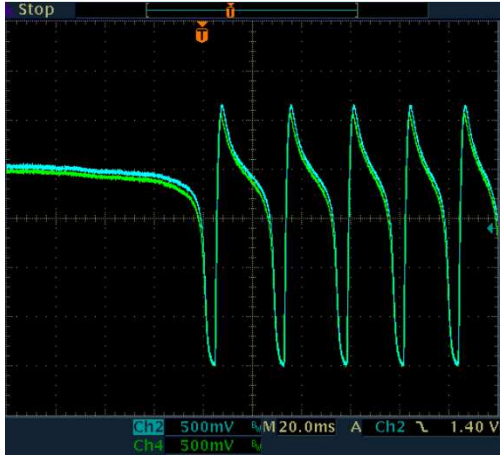}
\caption{\label{HSS_to_IP}Time series demonstrates HSS to IPLC transition in circuit for decreasing Q at 0.85 (LP1); $n=4, \alpha=182, \kappa=4.2$.}
\end{figure}

Increasing Hill repression beyond 3.2 eventually leads to the appearance of stable in-phase (IPLC) oscillations--a new dynamic not seen at the smaller $n$-values inside broad intervals of $\alpha$ and $\kappa$.  The $n$-continuation diagram in Fig.\ \ref{n-contink4} shows the APLC and IPLC, and the HSS for activation rate $\kappa=4, Q=0.8$, and $\alpha=175$.  Stable IPLC occurs in a restricted range of parameters $n, \kappa$, and Q.  For $\kappa=4, Q=0.8$, the range of repression for stable IPLC is $n= 4.12$ to 4.37.  As coupling strength Q is increased (decreased), the $n$-range for stable IPLC moves to lower (higher) $n$-values (results not shown).  Then, at a fixed $n$-value there is only a narrow Q-range of stable IPLC.  For example, at the higher value of $Q=1.0$ the stable IPLC occurs from $n=3.71$ to 3.89 with no overlap of the $n$-range for $Q=0.8$, and at the lower $Q=0.7$ the stable IPLC occurs from $n=4.36$ to 4.8 with overlap only at the upper edge of the $n$-range for $Q=0.8$.  The limited Q-range of stable IPLC is apparent in the Q-continuation diagram in Fig.\ \ref{n4a175k4} for $n=4, \kappa=4$.  (A linear vertical scale is used in Fig.\ \ref{n4a175k4} in order to make clear the APLC and IPLC branches.) As Q increases, the IPLC branch ends with an infinite period bifurcation transition to the LP1 of the HSS.  Figure \ref{HSS_to_IP} shows a screen-shot of the opposite transition, from HSS to IPLC, which occurs in the circuit when Q is slowly decreased through the LP1 of the high-B-HSS.  Note that the amplitude of the IPLC is larger than the HSS as predicted in Fig.\ \ref{n4a175k4}.

It was shown above that increasing the rate of activation $\kappa$ causes the LP1 of the HSS to move to lower Q-values.  This decrease must then also move the end-point of the IPLC branch to lower Q-values, which can result in loss of the stable IPLC.  For example, raising $\kappa$ to 5 with Q=0.8 causes the $n$-range for stable IPLC to move to $n=3.8$ and shrink to only 0.008 width (compared to width 0.25 at $n=4.25$ for $\kappa=4$ in Fig.\ \ref{n-contink4}), and it causes the complete loss of stable IPLC for all $n$-values for $Q > 0.8$.  The Q-interval for stable IPLC can be made longer by increasing $\alpha$ (data not shown). Appearance of the stable IPLC is the only new stable attractor at the larger cooperativity of repression. The IPLC coexists with stable APLC, which remains the dominant dynamical behavior. 

\subsection{Unstable APLC Regime, Quasiperiodicity, and Chaos}
\begin{figure}
\begin{subfigure}{0.495\textwidth}
\centering
\includegraphics[width=0.85\linewidth]{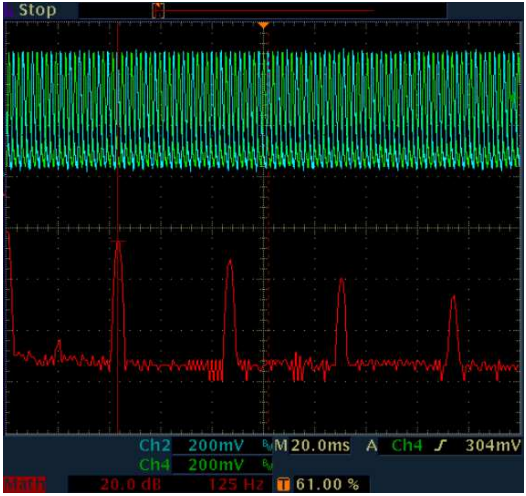}
\caption{\label{fft-AP}AP and FFT. $Q=0.6$.}  
\end{subfigure}\
\begin{subfigure}{0.495\textwidth}
\centering
\includegraphics[width=0.85\linewidth]{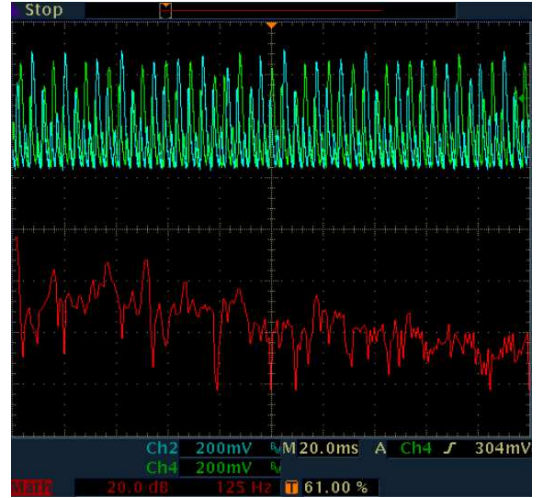}
\caption{\label{fft_co}CO and FFT. $Q=0.9$}  
\end{subfigure}
\caption{\label{n3scrnshots}Oscilloscope screenshots of time-series and FFT. $n=3.0,\alpha=190, \kappa=17.8$.}
\end{figure}

In the region of unstable APLC between the TRs (see Fig.\ \ref{n3a190k15}) a rich variety of behaviors are predicted numerically and observed in the circuit for $n \approx 3.0-3.4$.  Figure \ref{n3scrnshots} shows screenshots of simple APLC and complex oscillations (CO) in the circuit for the Hill coefficient $n= 3.0$ and coupling strengths Q=0.6 and 0.9.  The nature of the dynamics in the circuit was investigated by using the oscilloscope's built-in fast fourier transform (FFT) function.  Figure \ref{fft-AP} for $n=3, \kappa=17.8, Q=0.6$ shows APLC oscillations in the upper portion which are stable as indicated by distinct peaks in the FFT in the lower portion. The DC component peak is seen at the left edge and peaks for the fundamental (at 270 Hz) and the 2nd, 3rd, and 4th harmonics are clearly apparent.  Figure \ref{fft_co} for the higher coupling $Q = 0.9$ shows irregular oscillations in the upper portion and a continuum FFT in the lower portion which suggests that oscillations are possibly chaotic.  The CO begin with the appearance of sub-harmonic peaks in the FFT for Q just beyond TR1 bifurcation ($Q=0.8$). The continuum FFT predominates for intermediate Q values, followed by the return of sub-harmonic peaks when Q approaches TR2 ($Q=1.1$) where a transition to the small-amplitude APLC occurs. Through this long interval of coupling strengths various transformations of CO types take place depending on the values of $\alpha$. CO can lose stability within certain Q ranges whereas clear evidence of chaotization are observed in others. The detailed study of CO evolution is not our goal in this work and we use this designation as a general term, keeping the term ``chaos'' only for when we are sure about the chaotic character of the dynamics. In circuits no transitions to HSS were observed for Q values varied gradually between the TRs for $n=3.0$.

\begin{figure}
\begin{subfigure}{0.495\textwidth}
\centering
\includegraphics[width=0.85\linewidth]{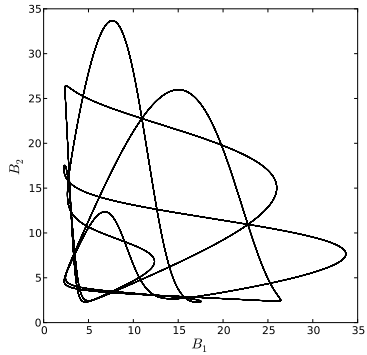}
\caption{\label{P5}5:5 limit cycle.}  
\end{subfigure}\
\begin{subfigure}{0.495\textwidth}
\centering
\includegraphics[width=0.85\linewidth]{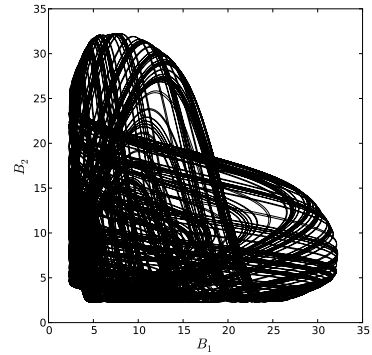}
\caption{\label{CO}Complex oscillations.}
\end{subfigure}
\caption{\label{phsplts}Phase plots of coexisting dynamics. $n=3.0,\alpha=200, \kappa=15, Q=0.935$.}
\end{figure}

Simulations with $n=3.0, \alpha=190, \kappa=15$, and Q ranging from 0.75 to 1.10 reveal that CO are the dominant behavior between the TRs and that there are no transitions to HSS, even with the addition of intensive white noise. The CO regime persists, as observed in the circuit although there may be differences in the nature of the CO attractor depending on the values of coupling strength and $\alpha$. To study the background of the CO regime for $n=3$ we use simulations to avoid the influence of possible circuit noise and parameter mismatches. The main dynamical surprise found between TRs over a large range of parameter $\alpha$ (180 to 300) is the coexistence of the CO regime and periodic limit cycle with five subperiods (``return times'') labeled below as 5:5LC. Figure \ref{phsplts} presents the phase portraits of 5:5LC and CO calculated using identical parameters. It is clearly seen that 5:5LC and this type of CO are anti-phase spatially symmetric attractors similar in structure and slightly different in amplitudes. 

\begin{figure}
\includegraphics[width=0.5\textwidth]{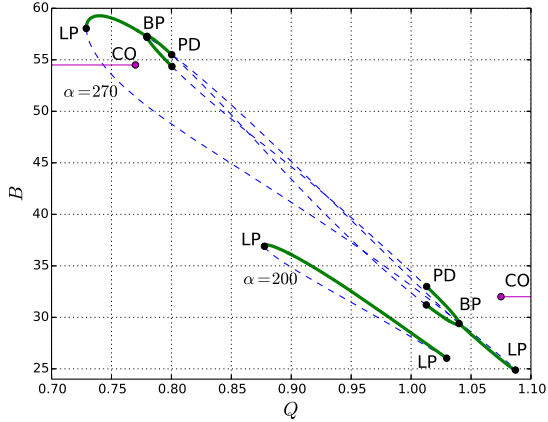}
\caption{\label{n3Qbif_5:5}Q-continuations of 5:5LC, solid (dashed) line indicates stable (unstable) attractors for $\alpha =200$ and 270, $n=3, \kappa=15$. Complex oscillations (CO) coexist with entire 5:5LC Q-interval for $\alpha=200$. Violet lines ending at small circles indicate regions of CO for $\alpha=270$. BP--branch point (broken symmetry bifurcation), PD--period doubling bifurcation, LP--limit point.}
\end{figure}

\begin{figure}
\begin{subfigure}{0.495\textwidth}
\centering
\includegraphics[width=0.95\linewidth]{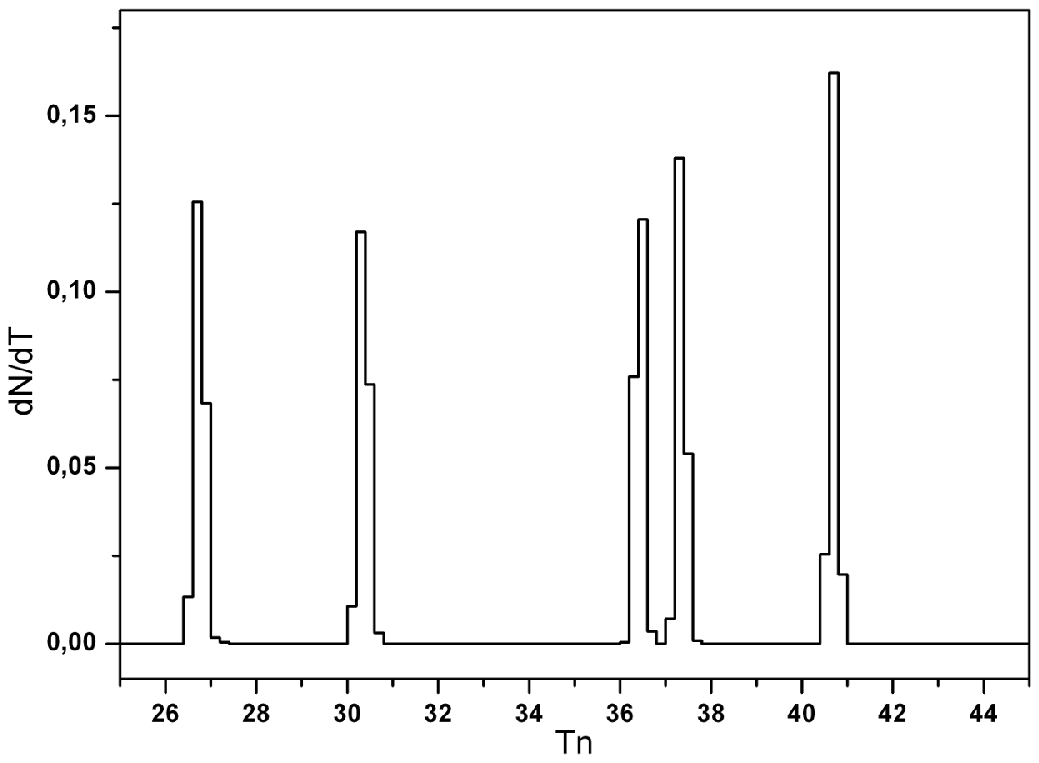}
\caption{\label{dNdt-P5}5:5 limit cycle.}  
\end{subfigure}\
\begin{subfigure}{0.495\textwidth}
\centering
\includegraphics[width=0.95\linewidth]{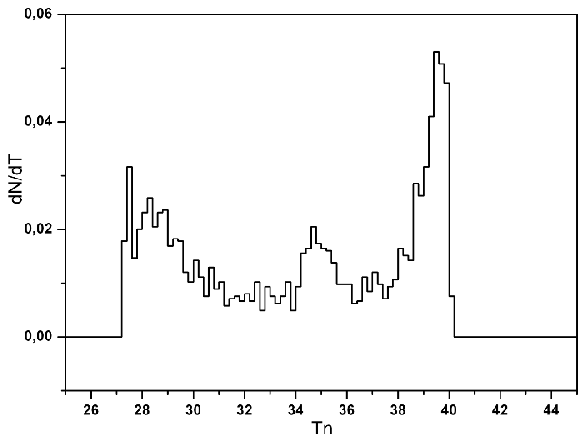}
\caption{\label{dNdt-CO}Complex oscillations.}
\end{subfigure}
\caption{\label{dNdt}Distributions of return times collected from the Poincar\'{e} sections of trajectories of 5:5LC and CO in the presence of small noise. $n=3.0,\alpha=200, \kappa=15, Q=0.935$.}
\end{figure}

Figure \ref{n3Qbif_5:5} shows two qualitatively different one-parameter continuations of 5:5LC which demonstrate the basic bifurcations controlling the dynamics of the system. For a particular set of parameters ($\alpha =200, \kappa=15$) the CO regime occupies the entire Q-region between TR bifurcations at 0.74 and 1.14 (not shown) thereby coexisting with the entire 5:5LC bounded by LP bifurcations at 0.879 and 1.03. The CO demonstrate quasiperiodic behavior after the lower TR up to the lower LP where stable 5:5LC emerges.  In the region of their coexistence narrow periodic windows appear in the CO indicating a change in the CO dynamics.  Figure \ref{dNdt} presents the distributions of return times of the coexisting attractors, calculated using Poincar\'{e} sections ($B_1=B_2=5$) of very long trajectories in the presence of uncorrelated low-level white noise added to variables $B_i$. Integrations were started from initial conditions corresponding to 5:5LC (Fig.\ \ref{dNdt-P5}) or to CO (Fig.\ \ref{dNdt-CO}).  Noise amplitudes were varied to find the most appropriate value to escape the endless and fast mixing of dynamic regimes. There is an interval of noise amplitudes which stimulates the appearance of visible dispersion of all five peaks (corresponding to the return times in Fig.\ \ref{dNdt-P5}) in dN/dT distributions, but it is too small to stimulate switching between the periodic 5:5LC and the CO. This means that despite similarities in the extent of phase relations (Fig.\ \ref{phsplts}) and in the boundaries of return times (Fig.\ \ref{dNdt}), the system Eq.\ \eqref{ode} has two types of coexisting stable solutions--5:5LC and CO--over large intervals of parameters inside the region of unstable APLC.
 
Figure \ref{n3Qbif_5:5} shows the effect increasing $\alpha$ has on the coexisting 5:5LC and CO attractors.  At $\alpha=200$ these attractors coexist for the entire Q-range between the LPs of the 5:5LC, however at $\alpha=270$ both attractors have lost continuity. The 5:5LC has a PD-cascade to chaos just beyond the lower PD, and a period-halving cascade back to stable As5:5LC at the upper PD, and then 5:5LC at BP. The regions of stable CO for $\alpha=270$ are marked by violet lines ending in circles, resulting in coexistence with 5:5LC just near the LPs. Chaos born from PD-cascades is the sole oscillating attractor over the broad Q-range between PDs. For $\alpha<295$ the chaos attractor remains separated from the CO attractor by Q-regions where As5:5/5:5LC is the sole oscillating attractor. 

A more in depth examination of the dynamics for $n=3.0$ is in SM. Figure SM5 shows additional detail of the evolution of the 5:5LC as $\alpha$ increases. Figure SM6 shows a map of regimes in $(Q,\alpha)$-space clearly showing coexistence of CO and 5:5LC, and the separation of CO and chaos emerged from PD-cascades. Figure SM7 shows an As5:5LC measured from a circuit.    

In summary, raising the Hill coefficient to $n=3$ opens new complex dynamic behaviors which are different from classic chaos found earlier for two coupled repressilators \cite{ullner2007,ullner2008}: the coexistence of the CO and 5:5LC attractors for $\alpha \approx 200$; then for increasing $\alpha$, the loss of stable CO over a Q-range, followed at higher $\alpha$ by the loss of stable 5:5LC over a Q-range and its replacement by chaos born from PD-cascades.  The chaos attractor is distinct in character from the CO attractor, and they are separated by Q-regions where 5:5/As5:5LC is the sole oscillating attractor over a large $\alpha$-parameter range.  For the investigated broad areas of model parameters 5:5LC is ``isola'' in terms of bifurcation theory, not linked to other regimes in the investigated ranges of parameters. Bifurcation analysis of its evolution as a function of Q and $\alpha$, its competition with CO, as well as its transformation into chaos in parameter space deserves special investigations beyond the scope of this paper (work in progress).

\begin{figure}
\begin{subfigure}{0.485\textwidth}
\centering
\includegraphics[width=0.6\linewidth]{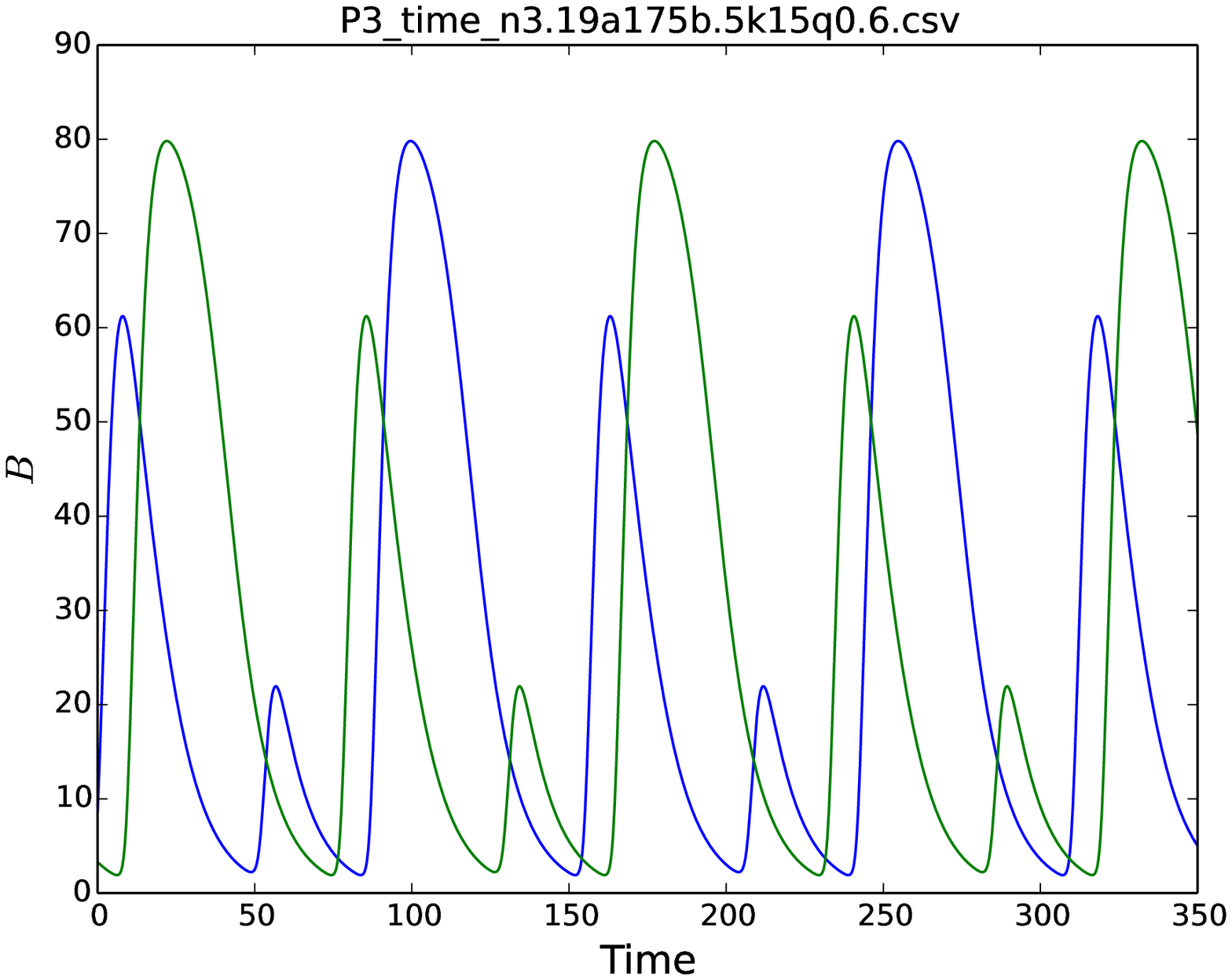}
\caption{\label{P3-num}3:3LC numerical. $n=3.19$, $\kappa=15$, $Q=0.6$.}  
\end{subfigure}
\begin{subfigure}{0.485\textwidth}
\centering
\includegraphics[width=0.6\linewidth]{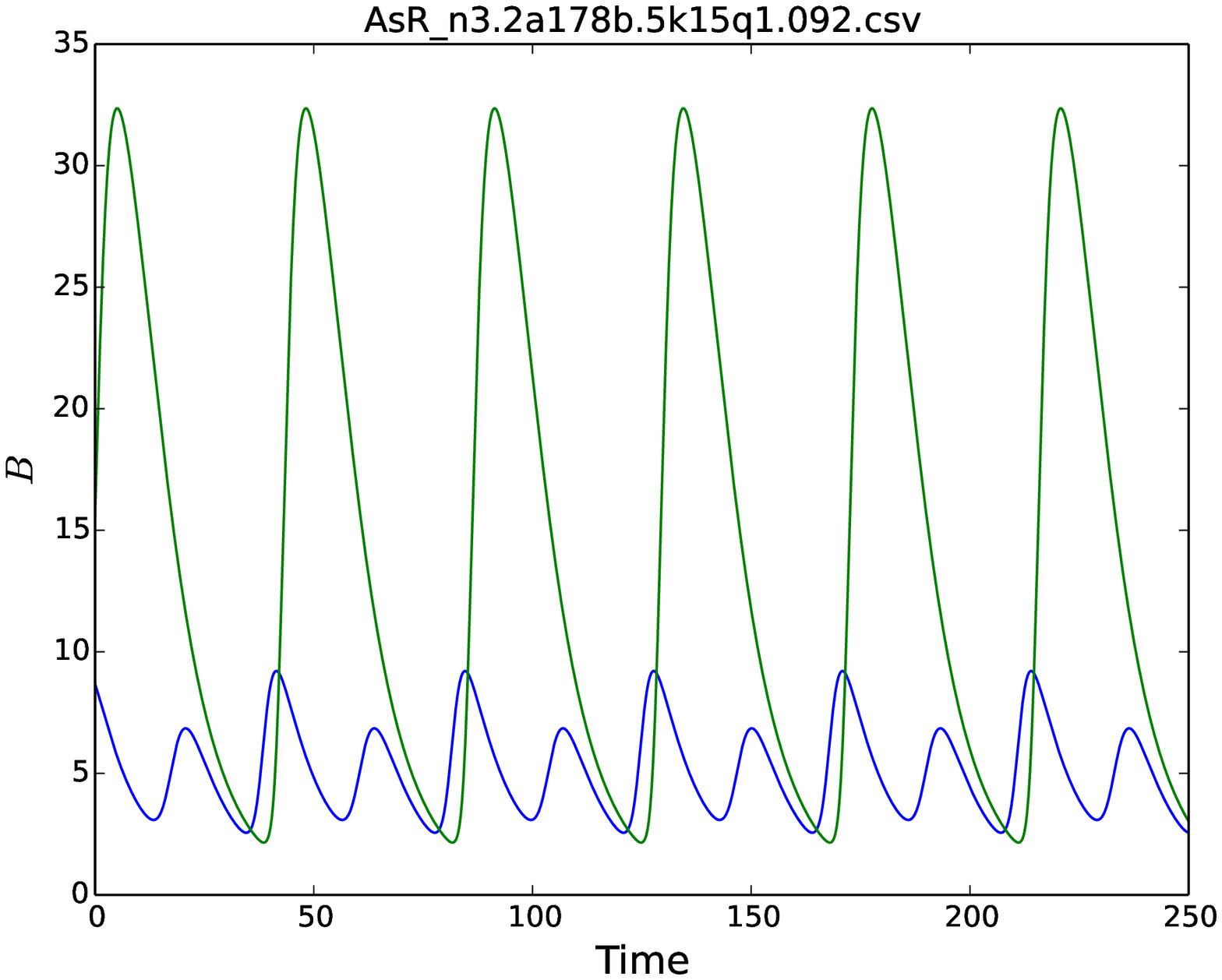}
\caption{\label{As1-2-num}As1:2LC numerical. $n=3.2$, $\kappa=15$, $Q=1.092$.}  
\end{subfigure}
\\
\begin{subfigure}{0.485\textwidth}
\centering
\includegraphics[width=0.6\linewidth]{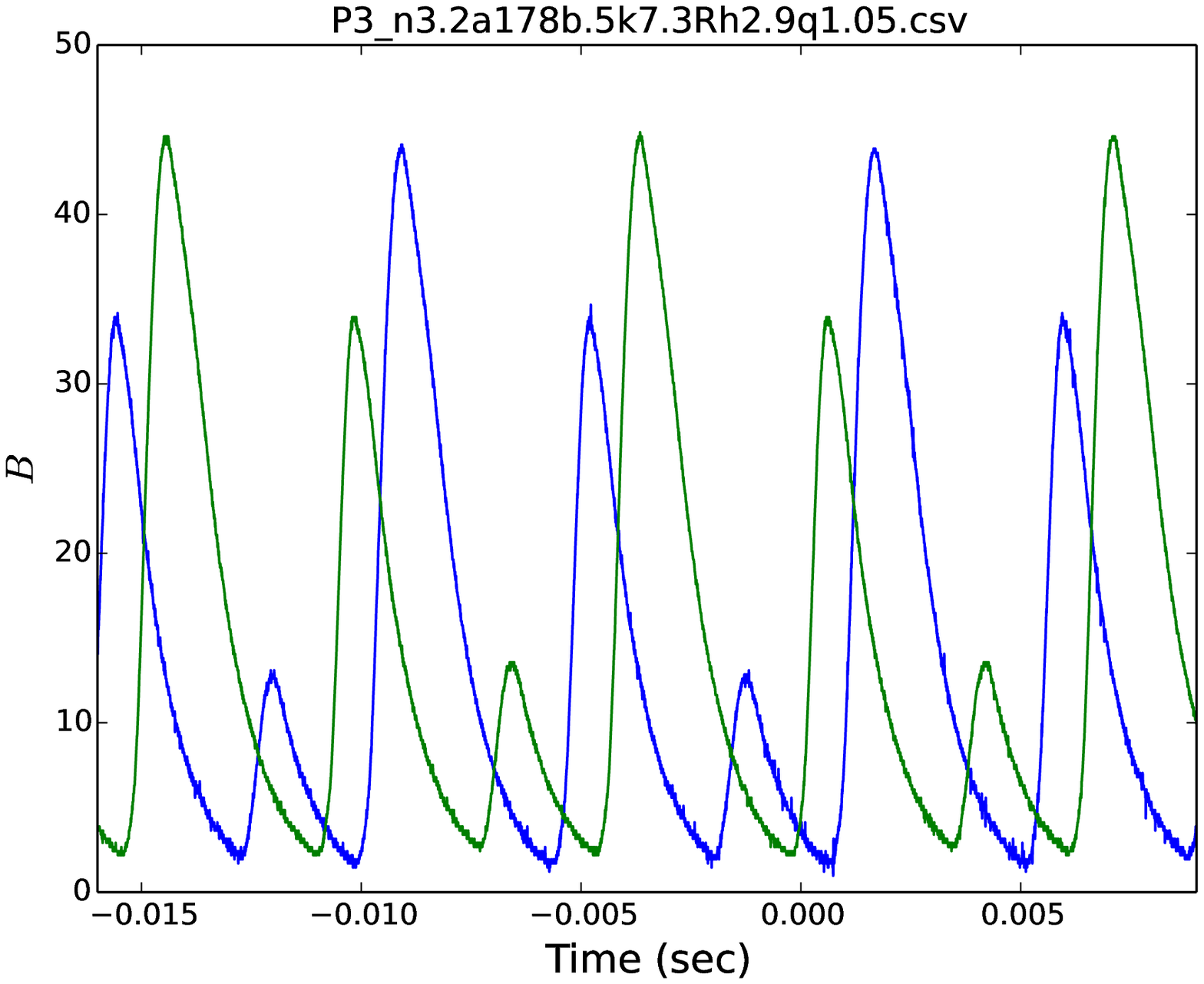}
\caption{\label{P3-exp}3:3LC measured. $n=3.2$, $\kappa=7.3$, $Q=1.05$.}  
\end{subfigure}
\begin{subfigure}{0.485\textwidth}
\centering
\includegraphics[width=0.6\linewidth]{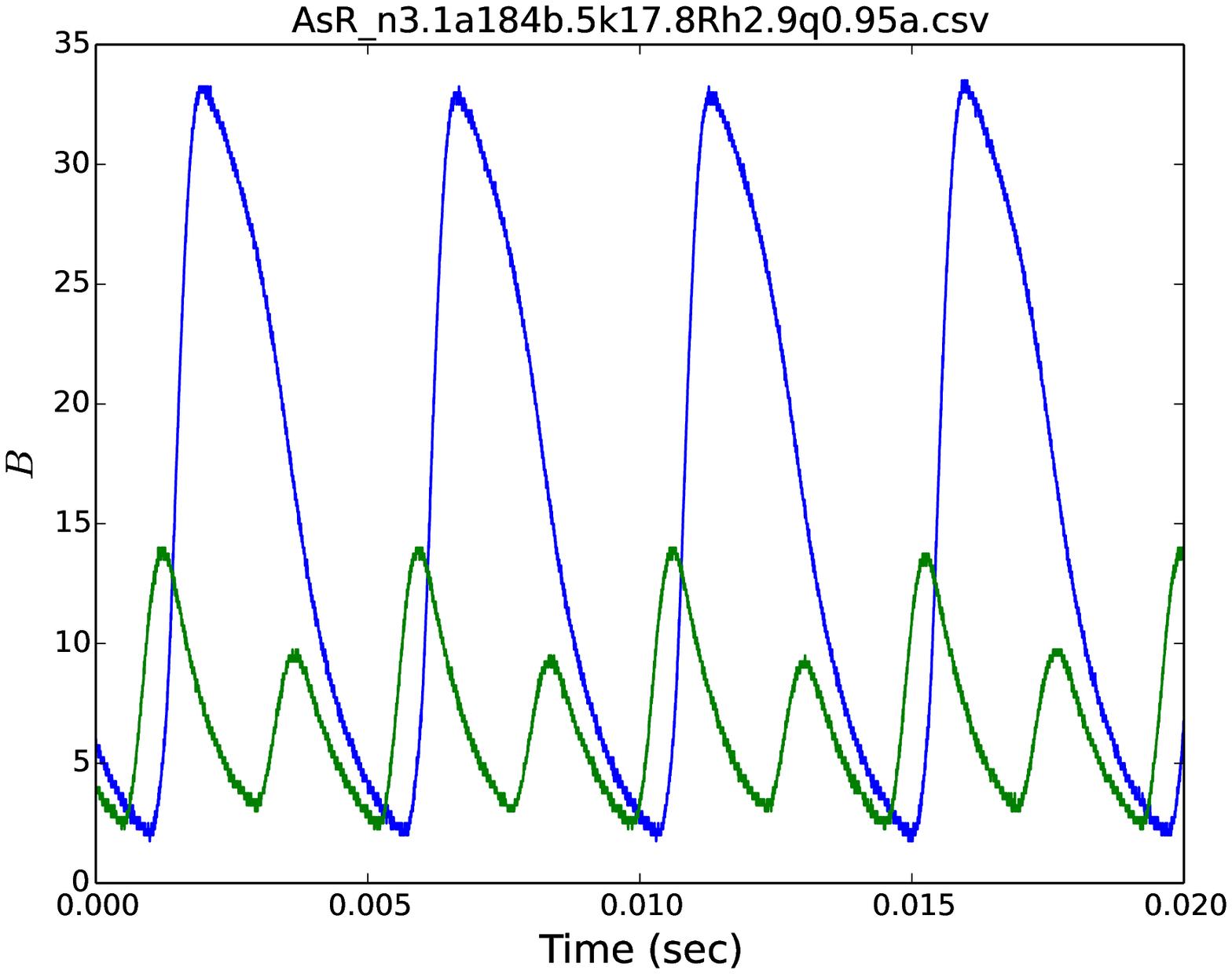}
\caption{\label{As1-2-exp}As1:2LC measured. $V_{th}=16$ mV and $R_{hill}=2.9$ $k\Omega$.}  
\end{subfigure}
\caption{\label{As}Plots of numerical simulations and measured limit cycles.}
\end{figure}

Next we consider a higher repression, $n = 3.15$ to 3.2, which leads to new qualitative changes in the dynamics in the region between TR bifurcations despite the small increase in the Hill coefficient of cooperativity.  By way of example, we show time series of some new dynamics from both simulations and circuit measurements, then we describe in detail the Q-continuation progression of the unstable APLC-branch dynamics for a representative case of $n=3.15$, $\alpha =225$, and $\kappa =10$.  Additional examples of behaviors from simulation and circuit measurement are shown in Supplementary Materials (SM).  

The basic regular attractors between TR bifurcations are high-period symmetric limit cycles n:nLC and a set of spatially asymmetric limit cycles (AsLC) with different oscillation amplitudes. Numerical simulations of stable 3:3LC and As1:2LC behaviors are shown in Figs.\ \ref{P3-num} and \ref{As1-2-num}, and their observations in circuit are shown in Figs.\ \ref{P3-exp} and \ref{As1-2-exp}.  In the circuit we search for unstable APLC behaviors by varying component values, with the understanding that no set of values gives a perfect match to the Hill function repression for a particular value of $n$.  Therefore when looking for similar behaviors in simulation and circuit measurement as in Fig.\ \ref{As} we are not concerned with exact matches of parameter values ($n,\alpha,\kappa$) between circuit and simulation.  
 
\begin{figure}
\includegraphics[width=0.5\textwidth]{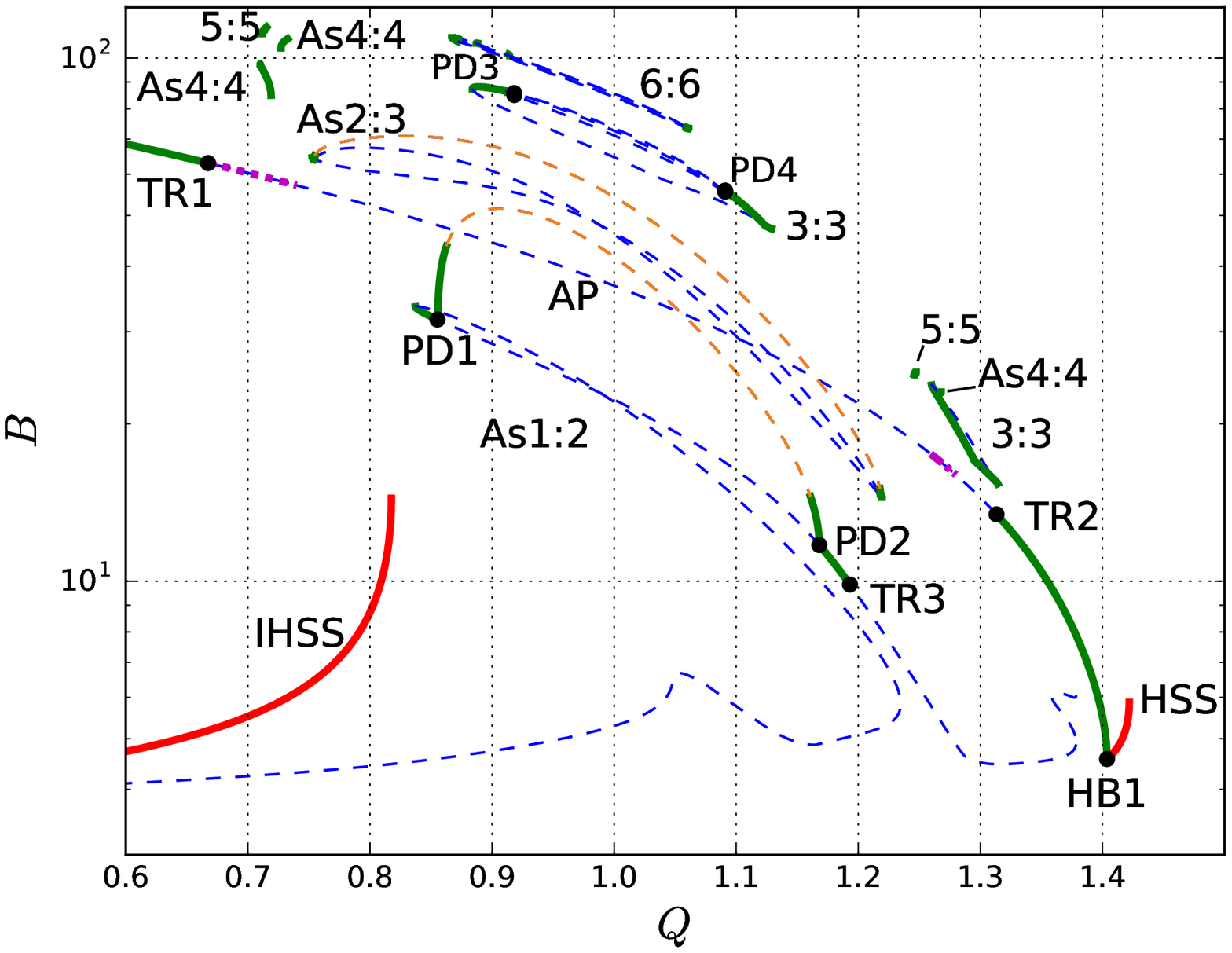}
\caption{\label{n3pt15}Q-continuations of stable (green) and unstable (broken lines) LC, and stable SS (red) for $n=3.15$, $\alpha=225$, $\kappa=10$. Narrower windows of stable high period LC not shown. For clarity only the lower amplitude oscillation of asymmetric LCs is shown.  The first period-doubled branches for As1:2 and As2:3 are shown (stable--green, unstable--broken orange). Regions of ``simple'' anti-phase complex oscillations (broken violet).}  
\end{figure}

Figure \ref{n3pt15} shows the Q-continuation of several attractors over the region where APLC is unstable. The basic regime branches---stable APLC, IHSS, and HSS---are familiar from previous diagrams as are the torus bifurcations TR1 and TR2, and the Hopf bifurcation HB1. However, Fig.\ \ref{n3pt15} shows many new regimes in multistable dynamics of system Eq.\ \eqref{ode}: high period limit cycles for 3:3LC (similar to Fig.\ \ref{P3-num} for $n=3.19$), 6:6LC, and spatially asymmetric limit cycles (similar to Fig.\ \ref{As1-2-num}).  Unstable portions are shown with broken lines for clarity. Four of the period doubling points are shown.  Also shown are five narrow windows with stable attractors: As4:4LC from LP(Q=0.710) to LP(Q=0.719), the stable ends of a 5:5LC at LP(Q=0.712) and LP(Q=1.247), and the stable ends of an As4:4LC at LP(Q=0.727) and LP(Q=1.268) (very close to 3:3LC). The unstable portions of the 5:5LC and the As4:4LC are not shown in Fig.\ \ref{n3pt15} due to confusing overlap with other branches.  (The 5:5LC may be the evolution of the 5:5LC shown above for $n=3.0$.) Many other narrow windows with stable symmetric and asymmetric high-period LCs also exist but are not shown. The numerous coexisting dynamical branches have a strong influence on the nature of the CO inside the region of unstable APLC. We use Fig.\ \ref{n3pt15} as a reference guide as we describe simulation results found using fixed values of coupling strength Q.  Note that Fig.\ \ref{n3pt15} zooms in to make clear the various dynamic branches, and therefore does not show the high-B-HSS which is apparent in previous Figs.\ \ref{n2.8k12} and \ref{n3a190k15}.  The stable high-B-HSS competes with the other attractors for phase space over the entire region of unstable APLC.

Similar to the case of $n=3$, the first regime after the TR1 bifurcation is a quasiperiodic attractor characterized by ``beating'' of the APLC time series producing two peaks in return times distributions. With further increase of Q, a third peak emerges suggesting the approach of chaos (all within the broken violet line in Fig.\ \ref{n3pt15} from TR1 to 0.735).  The more effective indicator of chaos is the emergence of a strongly folded sequential period map presented in Fig.\ SM8 to demonstrate the process of chaos maturation as Q grows.  

Within the region between TR1 and $Q=0.735$ there are many very narrow periodic windows which may be occupied by amplitude-symmetrical LCs (e.g. 13:13LC, 35:35LC and other n:n-LCs) or by slightly amplitude-asymmetrical regimes (e.g.\ the As4:4LC).  Many of these regimes in narrow windows demonstrate period doubling bifurcations returning system to chaos.  For example, the 5:5LC starts at 0.7116 as a stable symmetric 5:5LC which branches to slightly asymmetric 5:5LC at 0.7141 then starts a PD cascade to chaos at 0.7152. This regime coexists with asymmetrical 4:4LC (see their phase portraits and zoom-in Q-continuations in Figs.\ SM10 and SM11).  In contrast to the 5:5LC, the As4:4LC from Q = 0.710 to 0.719 is ``isola'' because its Q-interval is limited by LP bifurcations. The chaos emerged through the region of TR1 to 0.735 is characterized as ``simple'' anti-phase attractor similar to that presented above in Fig.\ \ref{CO} for $n=3$, which means that elements of the highly asymmetric behavior arise only for $Q > 0.74$. 

\begin{figure}
\begin{subfigure}{0.495\textwidth}
\centering
\includegraphics[width=0.95\linewidth]{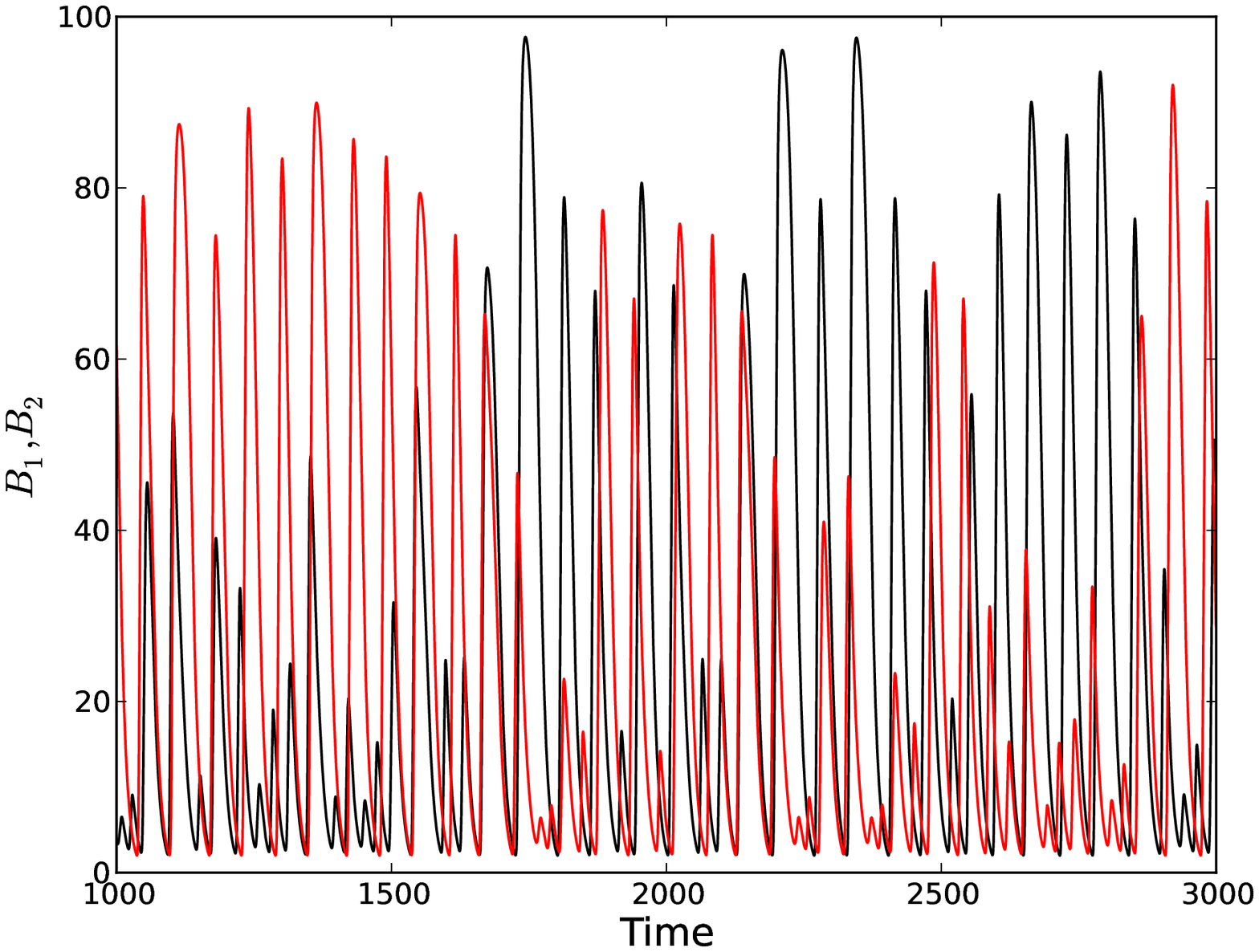}
\caption{\label{switch-chaos}Q=0.95.}  
\end{subfigure}\
\begin{subfigure}{0.495\textwidth}
\centering
\includegraphics[width=0.95\linewidth]{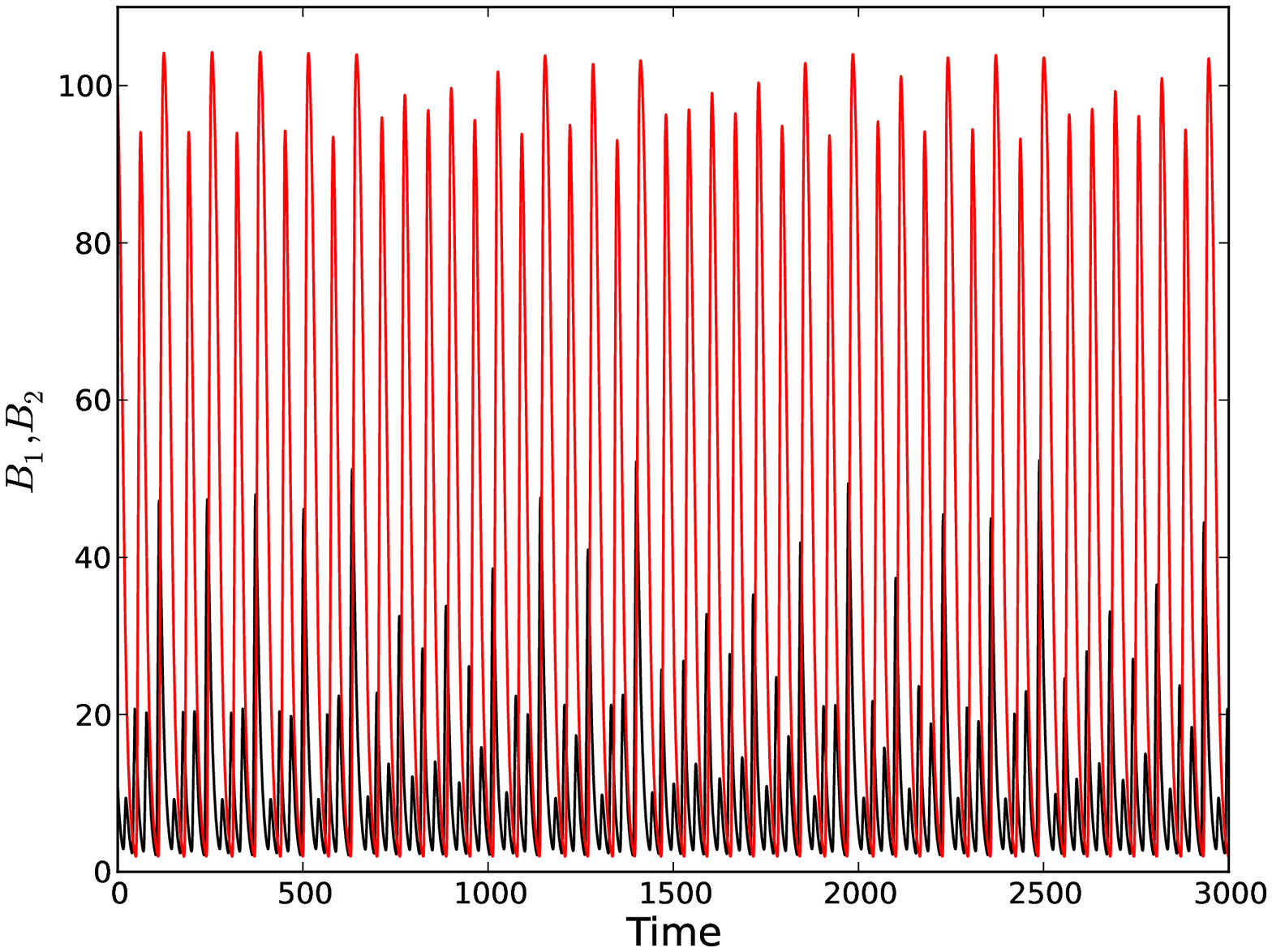}
\caption{\label{As-chaos}Q=0.87.}
\end{subfigure}
\caption{\label{chaos-types}The time series $B_1, B_2$ of chaotic attractor for $n=3.15, \alpha=225, \kappa=10$. Panel (a) shows branch switching resulting in symmetric chaos, (b) shows As-chaos as a result of no switching.}
\end{figure}

Spontaneous durations of large amplitude asymmetric rhythmic behavior, which is our main target here, begin to appear within chaos for Q beyond 0.74 thus ending the region of simple chaos.  These AsLC durations are related to the appearance of AsLC branches extending over a broad Q-range which have stable regimes near their low-Q and high-Q ends.  For example, Fig.\ \ref{n3pt15} shows continuations As2:3LC (low-$LP= 0.7528$) and As1:2LC (low-$LP= 0.837$) extending to the high-LP (1.219 for As2:3LC) and to TR3 (1.193 for As1:2LC).  Various stable n:n regimes also appear including portions of the 6:6LC, 3:3LC, and 5:5LC.  In contrast to the dynamics for $n=3.0$, there are no broad Q-intervals with the coexistence of chaos and stable limit cycles. The stable regimes (for both AsLC and n:n-LC) typically undergo period-doubling cascades to chaos at the low-Q end of their branch, and period-halving cascades (for increasing Q) from chaos back to the stable high-Q portion of the branch. The coexistence of numerous dynamical branches, each with intervals of stability and cascades to chaos, has a dramatic effect on the types of chaos manifestations.  Much of the chaos between Q = 0.74 and 1.25 consists of switching between pieces of the various coexisting unstable branches.  In particular, many pieces of As1:2LC are found in the chaotic trajectory over the Q-interval spanned by the As1:2-LC branch.  A typical example of time series observed for value $Q=0.95$, which is chosen far from the boundaries of periodic windows to exclude the overlay with intermittency, is presented in Fig.\ \ref{switch-chaos} (other examples shown in Figs.\ SM9 and SM12). It contains the alternating parts of asymmetric attractors as well as the elements of 3:3LC. 

The stable As1:2LC and As2:3LC both convert to purely asymmetric chaos (As-chaos) via period-doubling cascades (PD points for As2:3LC are not shown since they would obscure the small regions of stable LC). The first PD branches of As1:2LC and As2:3LC are shown using orange broken lines for their unstable portions. The stable As1:2LC extends from a limit point at 0.837 to PD1(0.855) and from PD2(1.168) to TR3(1.193). The period-doubled stable As2:4LC extends from PD1 to 0.8627 and from 1.1609 to PD2.  The As-chaos (located near the ends of the unstable PD branches in orange) exists over a narrow Q-interval beyond which it typically converts to the chaos consisting of switching between pieces of coexisting unstable branch dynamics.  The largest regime of purely asymmetric chaos (with no switching) occurs at the end of the PD1 cascade.  This cascade (As2:4LC at 0.855, As4:8LC at 0.8627) of As1:2LC ends in asymmetric chaos (Fig.\ \ref{As-chaos}) which coexists with the stable 6:6LC, with both the asymmetric chaos and the stable 6:6LC ending at Q=0.872.  The asymmetric chaos is distinct from the chaos composed of pieces of asymmetric behavior, which over time is symmetric due to the chaotic switching of oscillator’s amplitudes (Fig.\ \ref{switch-chaos}).

The existence of different nearby stable attractors determines the creation of intermittent behaviors composed from different combinations of unstable attractors. Typical examples of intermittency localized near the boundaries of periodic windows are exhibited in Figs.\ SM14-SM17 but their analysis is beyond the scope of this paper. 

Beyond $Q \approx 1.25$ the number of coexisting dynamical branches is reduced, causing the chaotic behavior to change character (indicated by the broken violet line in Fig.\ \ref{n3pt15}). The chaos is characterized by pieces of symmetric n:n-LC behavior and contains none of the highly asymmetric behavior associated with the As1:2LC and As2:3LC branches. (Apparently the unstable As1:2LC portion beyond TR3 does not contribute to the chaotic behavior.)  The chaos coexists with the stable 3:3LC out to about 1.28.  For Q from 1.28 to TR2 the stable 3:3LC and the HSS capture all the behavior.  ``Beating'' exists just below TR2, occurring at Q = 1.3104 and converting to 3:3LC at Q=1.3102, and is mentioned here only to show the qualitative similarity of dynamics near TR1 and TR2 bifurcations. 

Finally, we consider the unstable APLC region for $n$ beyond 3.2.  Generally, simulations find that parameters $\alpha$ or $\kappa$ have to be changed in order to find interesting regimes (chaos, asymmetric limit cycles, symmetric regular n:n-LC, intermittency) like those demonstrated for $n$ around 3.15.  For example, for $n=3.3$ and $\kappa=15$ the interesting regimes appear only if $\alpha<120$.  In contrast, for $\kappa=4$ the system demonstrates only stable APLC for $100<\alpha<200$, however the typical complex evolution $Q(TR1)\rightarrow 5:5 \rightarrow 4:4 \rightarrow 1:2 \rightarrow 3:3$ is restored when $\alpha=250$ with $Q>0.945$. Hill coefficient $n=3.5$ with $\kappa=15$ requires further reduction in $\alpha$ (55-70) for the existence of interesting multistability which are then located at $0.7<Q<1.1$.  If activation is significantly reduced ($\kappa=4$) then $\alpha$ must be increased to get interesting regimes.  However, these regimes are then located at $Q>1$ if $100<\alpha<180$.  We conclude that limited but significant variations of basic parameters in 3-dim space do not discover other qualitatively new dynamic regimes compared to those described above. 

\section{Discussion}

Using simple 4-dim mathematical and electronic models of identical ring oscillators we demonstrated that the realistic design of quorum sensing coupling proposed earlier \cite{ullner2008} leads to remarkable multistability even for a pair of oscillators. A QS-coupling system can be incorporated into genetic circuits in different ways, acting solely as a coupling agent (as in the case of repressilators \cite{garcia2004} and Relaxators \cite{mcmillen2002}), or it may also be an integral part of an individual oscillator \cite{kuznetsov2004,danino2010,prindle2014}. In any case, the additional differential equation for autoinducer contains a term describing its linear diffusion but the resulting coupling is nonlinear for the oscillators because of a delay in the autoinducer production and because the production of one of the oscillator's variables is activated nonlinearly. Therefore it is natural to expect unusual collective modes for QS-coupled oscillators, as was demonstrated in \cite{ullner2008}. However, that study of multistability considered mainly the role of coupling strength, with the other parameters of the single oscillator like repression cooperativity and the transcription rate being fixed. Therefore, the generality of QS-induced multistability was still obscure because the origin of many new regimes can be traced to modulations in oscillation amplitudes and the ``steepness'' of repression which are strongly dependent on the dynamic properties of the isolated oscillators.

We extended the study by addressing the roles of the other parameters and detected an interesting evolution of multistability in the 3-dim parametric space defined by Hill coefficient of repression ($n$), AI-induced activation ($\kappa$) and the rate of transcription ($\alpha$). This extension is not a pure theoretical exercise; rather, it is stimulated by experimental progress in the development of engineered promoters having different strengths and structures that control the cooperativity in binding of transcription factors.

Combination of numerical simulations, bifurcation analysis and electronic measurements is beneficial to investigations of dynamical systems like the synthetic genetic network studied here.  Benefits come from identifying the similarities and the differences between results of the numerical and electronic models.  Dynamic characteristics that are similar in both systems are unlikely to be particularly sensitive to the model parameters, and, therefore, could be present in other similarly constructed systems, possibly even real biological systems in the future. 

For small $n$ around 2.8 our reduced 4-dim repressilator models reproduce the basic regimes found in the 7-dim model used in \cite{ullner2008}: stable anti-phase limit cycle started for small and ended at high coupling strength and stable homogeneous and inhomogeneous steady states, as well as inhomogeneous limit cycle at intermediate coupling strengths. Multiple regimes may coexist over a broad range of control parameters. 

Further smooth increase in Hill repression $n$ in association with the respective adaptation of activation $\kappa$ introduces an unstable APLC via torus bifurcation over a region of coupling strengths as demonstrated by the results for $n=3.0$. With increasing coupling strength, the dynamics in the region where the APLC is unstable shows gradually developing complex oscillations (as can be detected using FFT analysis and/or from calculations of subperiod distributions), which begin with beating near the TR bifurcations. An extraordinary property of the dynamics for $n=3$ is the coexistence of the developing complex oscillations and the stable limit cycle consisting of five subperiods (see Figs.\ \ref{phsplts} and \ref{n3Qbif_5:5}) which itself is an isolated solution. The Q-intervals of coexistence strongly depend on the amplitudes of isolated repressilators, being very long for $\alpha=200-220$ and nearly negligible for $\alpha>350$.  Starting from intermediate $\alpha$ ($\alpha>250$ for our parameter set) the emergence of chaos at low Q and its extinction at higher Q are the result of sequential branch point and period doubling bifurcations of this 5:5LC. Each local bifurcation of our system on the $(\alpha, Q)$ plane for $n=3$ is well known but the structure of the map of regimes as a whole is unusual to the best of our knowledge. 

We found the interval of $n$-values where chaos in two QS-coupled repressilators is not an exotic regime but instead exists in a large 3-dim parametric space. Moreover, a further increase in $n$ up to $n=3.15$ introduces new behaviors inside the Q-interval with the unstable regimes, including regular and irregular asymmetrical ones (Figs.\ \ref{As} and \ref{chaos-types}). All collective modes presented in our study are the result of competition between repression (Fig.\ \ref{eHill}) and activation (Fig.\ \ref{QS-activation}) of transcription of one specifically selected gene. These antagonistic impacts have different dependencies on the values of the variables: repression is cooperative while activation is linear with saturation. This means that variation of one parameter, e.g.\ Hill repression $n$, may require the other parameters be tuned to escape the transition of the system to a simple attractor like HSS.

We investigated several sets of parameters and found that the evolutions of the complex regimes as a function of coupling strength are qualitatively very similar. Then we looked over the typical dynamics of chaos and discovered its basic skeleton although many details are still unclear. The most impressive result is the existence of several types of chaos: (1) simple anti-phase chaos riddled with very narrow periodic windows of symmetric regimes of the n:n-LC type; (2) asymmetric chaos resulting from period doubling of regular asymmetric regimes, e.g.\ As1:2LC and As2:3LC; and (3) symmetric chaos consisting of symmetric and asymmetric parts with alternating polarities. The last type of chaos is the most flexible in its specific manifestations and is closely linked to intermittency which is also observed but only near the boundaries between the regimes.

The possibility for chaos due to linear diffusion between \textit{identical} regular oscillators is known since the middle of the 1970s \cite{rossler1976,fujisaka1980,schreiber1982,hocker1989}, but the parameter regions of chaos are typically very narrow. Chaos via torus bifurcation of anti-phase limit cycle has been observed in two R\"{o}ssler oscillators coupled by cross-diffusion \cite{waller1984}. To the best of our knowledge there is only one example of chaotization due to nonlinear coupling of identical biochemical oscillators \cite{sporns1987} where ``bichaoticity'' was demonstrated in narrow parameter intervals without analysis of other regimes inside chaos.

A large cooperativity $n$ (around 4) opens the gate in parameter space for the in-phase limit cycle which becomes stable for large Q before its transition to HSS via infinite period bifurcation. It is still not clear why phase-repulsive coupling permits the formation of APLC/IPLC switching. Intuitively it may be due to the increase in repressilator ``stiffness'' for large values of $n$. Similar frequency trigger covering a wide range of parameters was observed in the system of two FitzHugh-Nagumo oscillators coupled via the recovery variable if their stiffness was large \cite{volkov1991}.  

All the results discussed above have been obtained for sets with different $\beta_i$. The oscillators are identical but move along the partial limit cycles nonuniformly because the kinetics of the specific variable in one element of the ring is faster than in the other elements. Variation in $\beta_i$ is biologically motivated given that there are parameters' differences between the three repressilator genes, their promoters, and degradation rates of the corresponding transcription factors. This internal variety in $\beta_i$ may enhance the coupling-induced diversity of the attractors. To check this possibility we simulated the dynamic behavior in the 3-dim parameter cube $(n, \alpha, \kappa)$ using several sets of $\beta_i$ varying in the degree of equalization (e.g. 0.25, 0.15, 0.15; 0.3, 0.25, 0.15) (data not shown). The only general effect of smoothing the $\beta_i$ is the necessity to tune reasonably the other parameters but the qualitative picture of the phase diagrams is not changed.

No doubt, the model in Eq.\ \eqref{ode} is a strongly reduced dynamic toy which has limited experimental support for formulation of exact forms of functions for the basic repressilators' ring and especially for the subsystem describing coupling. However, the general design formalized in Eq.\ \eqref{ode} is realistic and realizable by the methods of synthetic genetics. Therefore, we propose that a pair of coupled repressilators may be viewed as a prototype of a flexible generator with remarkable dynamic behavior which may be useful beyond the construction of synthetic genetic networks.    

\begin{acknowledgments}
This work was partially supported by grants RFBR 15-02-03236 and 14-01-00196 to E. V. 
\end{acknowledgments}

%
\end{document}